\documentclass[aps,onecolumn,showpacs]{revtex4}
\usepackage{graphicx}
\usepackage{amsmath,amscd,amssymb}
\usepackage{color}

\newcommand{\be}{\begin{equation}}
\newcommand{\ee}{\end{equation}}
\newcommand{\bea}{\begin{eqnarray}}
\newcommand{\eea}{\end{eqnarray}}
\newcommand{\lan}{\left\langle}
\newcommand{\ran}{\right\rangle}
\newcommand{\br}{\mathbf{r}}

\newcommand{\bq}{\mathbf{q}}

\newcommand{\bx}{\mathbf{x}}

\newcommand{\bk}{\mathbf{k}}
\newcommand{\e}{\varepsilon}

\newcommand{\eme}{\varepsilon_m}
\newcommand{\ew}{\varepsilon_w}
\newcommand{\tv}{\tilde{v}}
\newcommand{\tw}{\tilde{w}}
\newcommand{\tJv}{\mathbf{\tilde{J}_v}}
\newcommand{\tJ}{\tilde{\mathbf{J}}}
\newcommand{\tphi}{\tilde{\phi}}

\newcommand{\tbpsi}{\tilde{\mathbf{\varPsi}}}
\newcommand{\tbV}{\tilde{\mathbf{V}}}
\newcommand{\tbG}{\tilde{\mathbf{G}}}
\newcommand{\tpsi}{\tilde{\psi}}
\newcommand{\tG}{\tilde{G}}

\begin{document}

\title{Ion size effects at ionic exclusion from dielectric interfaces and slit nanopores}

\author{Sahin Buyukdagli$^{1}$\footnote{email:~\texttt{buyukds1@cc.hut.fi}}, C. V. Achim$^{1}$\footnote{email:~\texttt{cristian.achim@hut.fi}}, and T. Ala-Nissila$^{1,2}$\footnote{email:~\texttt{Tapio.Ala-Nissila@hut.fi}}}
\affiliation{$^{1}$Department of Applied Physics and COMP center of Excellence, Aalto University School of Science, P.O. Box 1100, FIN-00076 Aalto, Espoo, Finland\\
$^{2}$Department of Physics, Brown University, Providence, Box 1843, RI 02912-1843, U.S.A.}
\date{\today}

\begin{abstract}
A previously developed field-theoretic model [R.D. Coalson et al., J. Chem. Phys. \textbf{102}, 4584 (1995)] that treats core collisions and Coulomb interactions on the same footing is investigated in order to understand ion size effects on the partition of neutral and charged particles at planar interfaces and the ionic selectivity of slit nanopores. We introduce a variational scheme that can go beyond the mean-field (MF) regime and couple in a consistent way pore modified core interactions, steric effects, electrostatic solvation and image-charge forces, and surface charge induced electrostatic potential. Density profiles of neutral particles in contact with a neutral hard-wall, obtained from Monte-Carlo (MC) simulations are compared with the solution of mean-field and variational equations. A recently proposed random-phase approximation (RPA) method is tested as well. We show that in the dilute limit, the MF and the variational theories agree well with simulation results, in contrast to the RPA method. The partition of charged Yukawa particles at a neutral dielectric interface (e.g air--water or protein--water interface) is investigated. It is shown that as a result of the competition between core collisions that push the ions towards the surface, and repulsive solvation and image forces that exclude them from the interface, a concentration peak of finite size ions sets in close to the dielectric interface. This effect is amplified with increasing ion size and bulk concentration. An integral expression for the surface tension that accounts for excluded volume effects is computed and the decrease of the surface tension with increasing ion size is illustrated. We also characterize the role played by the ion size on the ionic selectivity of neutral slit nanopores. We show that the complex interplay between electrostatic forces, excluded volume effects induced by core collisions and steric effects leads to an unexpected reversal in the ionic selectivity of the pore with varying pore size: while large pores exhibits a higher conductivity for large ions, narrow pores exclude large ions more efficiently than small ones.

\end{abstract}

\pacs{03.50.De,05.70.Np,87.16.D-}
\maketitle

\section{Introduction}

Ion specific effects are relevant to an important number of biological, industrial and interfacial systems~\cite{Jung}. To mention a few examples, such effects are believed to play a crucial role in the absorption of large ions onto the water-air interface~\cite{Herce,Hagberg,NetzSim}, colloidal interactions in electrolyte solutions~\cite{Tavares}, charge reversal phenomenon~\cite{Molina}, protein stability~\cite{Cacace}, and the ionic selectivity of biological transmembrane channels that may discriminate between two different species of the same charge~\cite{Hodgkin,Hille,Bob}.

One of the earliest and probably the best known experiments that brings to light ion specific effects is the measurement of the positive surface tension of inorganic salt solutions by Heydweiller~\cite{heyd}. The observed enhancement of the surface tension of electrolytes compared to that of pure water was explained by Wagner~\cite{wagner} in terms of the screened image interactions of ions that lead to their depletion at the water--air interface. Within this theoretical picture, Onsager and Samaras obtained their celebrated limiting law that relates the surface tension of electrolytes to the bulk electrolyte density~\cite{onsager}. This limiting law is indeed able to reproduce experimental data for dilute electrolytes~\cite{Long}, but fails at large concentrations~\cite{Randles}. It is now well established that the discrepancy is due to the insensitivity of the Onsager-Samaras theory to ion specific effects, which are known to come into play at high densities. An earlier experimental observation of ion specific effects was made by Hofmeister who ranked salts according to their ability to precipitate egg white lysozyme~\cite{Hof}. It is known that there exists some correlation between Hofmeister series and the surface tension of electrolytes. Although it is believed that Hofmeister series are also correlated with the size and the intrinsic polarizability of ions~\cite{Pad}, a universal theory able to explain the chemical characteristics that characterize the series is still lacking. \textcolor{black}{Generalized electrolyte models taking into account at a MF level ion specific effects, such as ionic polarizability~\cite{PodgPol}, dielectric decrement~\cite{PodgDielDec} and ionic hydration~\cite{PodgHyd} have been developed.} Various theories taking into account steric effects~\cite{Levin,David} and ion polarizability~\cite{Bostrom} have been also proposed in order to improve on the Onsager-Samaras theory.

The finite ion size effects play also a key role in the case of electrolytes in contact with a strongly charged protein. Experimental evidence for such systems clearly shows ionic saturation effects close to the protein surface~\cite{Cuvillier}, which originate from close packing of ions. A modified Poisson-Boltzman theory that introduces steric effects by distributing the ions over a lattice have been proposed in Ref.~\cite{Boruk}. A different field-theoretic approach that includes excluded volume effects was also introduced in Ref.~\cite{Lue0}. Mean-field analysis of these theories was shown to successfully reproduce the saturation effect in question. However, these statistical models do not include a well-known  propriety of molecular liquids with repulsive interactions, namely the wetting of the interface by the fluid.

The structure of heterogeneous fluids has been a very active research domain over the past 50 years~\cite{Perkus1,Perkus2,HansBK} and early theories based on the solution of integral equations and mean-spherical approximations (MSA) were shown to reproduce with great accuracy the density profile of hard-core (HC) liquids in contact with an impenetrable wall~\cite{Henderson1,Lebowitz1}. Recent works have focused on the structure of heterogeneous liquids interacting with short-range square well potentials~\cite{HolovkoII},  Lennard-Jones~\cite{Zhou2} and HC-Yukawa potential mixtures~\cite{Zhou} within different perturbative density functional methods (DFT). The first analytical solution of the Ornstein-Zernike (OZ) equation for a purely repulsive bulk Yukawa liquid was derived within MSA  in Ref.~\cite{Waisman}.  Using a MF level DFT approach, it was recently shown in Ref.~\cite{Evans} that a binary heterogeneous fluid composed of repulsive Yukawa particles exhibits a wetting on the hard wall. Using a different method that consists in solving the OZ equation with the RPA closure, an analytical expression for the density profile of a repulsive Yukawa fluid was derived in Ref.~\cite{Holovko}. In contrast to Ref.~\cite{Evans}, the theory predicted a non-monotonic density shape, characterized by a depletion layer next to the wall, followed by a concentration peak and a decreasing particle concentration with increasing distance from the surface.

The coupling between electrostatic interactions and core collisions has been considered beyond the MF limit within integral equation theories. Using the reference hypernetted chain approximation, the role of various mechanisms, including ion size effects on the charge inversion phenomenon was analyzed in refs.~\cite{KjellanderI,KjellanderIII}. The authors investigated also the effect of image charge and ionic dispersion interactions on the interplate pressure and ionic partitions in Ref.~\cite{KjellanderII}. These integral equation methods developed within the particle density representation are efficient and provide good agreement with MC simulations~\cite{KjellanderIV}. However, they are also known to be computationally intensive and hard to use if one wishes to consider geometries other than planes. An alternative field theoretic model that couples electrostatic and core interactions was proposed in Ref.~\cite{DunYuk}. In this work, the void around an ion induced by the ion size was modeled with a repulsive Yukawa potential. By passing from the density to the field representation, the partition function of the system can be recast into a functional integral over two fields, namely a Yukawa field associated with core collisions and an electrostatic field induced by Coulomb interactions between charges. The authors considered the MF limit of this model in order to investigate the effect of ion size on the ionic partition next to two charged  spherical colloids as well as on the electrostatic interaction energy between the colloids. In the present work, we reconsider this field theoretic model within three computation schemes, among which a variational approach inspired by a previous variational method that has been developed for ions with purely Coulombic interactions at charged dielectric interfaces~\cite{PRE}. The variational formalism of Ref.~\cite{PRE} was also applied to the problem of ionic exclusion from cylindrical pores and yielded a new type of liquid-vapor phase transition that we proposed as the underlying mechanism behind the rapid switching of nanopore conductivity observed in experiments~\cite{PRL,JCP}.

This article is organized as follows. The derivation of the field theoretic model for charged Yukawa particles is briefly explained in Section~\ref{Mod}. Section~\ref{BulkLiq} deals with a bulk Yukawa liquid. Using a general variational ansatz, it is shown that at the first order variational level, short-range core interactions within the homogeneous medium experience a further screening but Debye-H\"{u}ckel interactions remain unchanged. In Section.~\ref{SlGe}, the physics of a heterogeneous Yukawa liquid is investigated. Various computation schemes are introduced and their validity for the density profile of neutral particles in contact with a hard wall is tested by comparison with MC simulations. The variational formalism is then applied to charged Yukawa particles in order to investigate core collision effects on the ionic distribution at a planar dielectric interface. An integral form for the surface tension taking into account the contribution from excluded volume effects is derived as well. In line with the Gibbs adsorption isotherm, core collisions that push the particles towards the interface lower the surface tension of the charged fluid. The variational equations are also solved in a slit pore in order to understand ion size effects on the ionic selectivity of slit nanopores. It is shown that due to a complex coupling between steric and excluded volume effects, image-charge forces and pore-modified screening of Coulomb and Yukawa interactions, large ions are more efficiently excluded from narrow pores than small ions while the penetration of large ions over small ions is favored into pores at nanometer scales. Limitations of the variational scheme, potential improvements and applications are discussed as well in the Conclusion.

\section{Model}
\label{Mod}

We present in this section the derivation of the field-theoretic partition function for charged Yukawa particles. Since the derivation was already introduced in Ref.~\cite{DunYuk}, we will exclusively report the general lines of the calculation. The canonical partition function of ions interacting with an electrostatic and a core potential reads
\be\label{PartCan}
Z_c=\prod_{i=1}^p\frac{e^{N_iE_s}}{N_i!\lambda_T^{3N_i}}\int\prod_{j=1}^{N_i}\mathrm{d}\bx_{ij}e^{-H_c\left(\{\bx_{ij}\}\right)
-H_y\left(\{\bx_{ij}\}\right)}
\ee
where $p$ is the number of ionic species, $N_i$ is the number of ions for each species and $\lambda_T$ is the thermal wavelength of an ion. The electrostatic and repulsive Yukawa interactions are respectively given by
\bea
\label{Hc}
H_{c}\left(\{\bx_{ij}\}\right)&=&\frac{1}{2}\int\mathrm{d}\br\mathrm{d}\br'\rho_c(\br)v_c(\br,\br')\rho_c(\br')\\
\label{Hp}
H_{y}\left(\{\bx_{ij}\}\right)&=&\frac{1}{2}\int\mathrm{d}\br\mathrm{d}\br'\rho_p(\br)w(\br,\br')\rho_p(\br')
+\int\mathrm{d}\br V_w(\br)\rho_p(\br)
\eea
with the charge density
\be
\rho_c(\br)=\sum_{i=1}^p\sum_{j=1}^{N_i}q_i\delta(\br-\bx_{ij})+\sigma_s(\br)
\ee
where $q_i$ stands for the valency of mobile ions, $\sigma_s(\br)$ is a fixed charge distribution, and the particle density is given by
\be
\rho_p(\br)=\sum_{i=1}^p\sum_{j=1}^{N_i}\delta(\br-\bx_{ij}).
\ee
The Coulomb and the short-range Yukawa potentials are defined respectively as the inverse of the following operators
\bea\label{coulomb}
v_c^{-1}(\br,\br')&=&-\frac{k_BT}{e^2}\nabla\left[\e(\br)\nabla\delta(\br-\br')\right]\\
\label{Yukawa}
w^{-1}(\br,\br')&=&\frac{b^2-\Delta}{4\pi\ell_y}\delta(\br-\br')
\eea
where $\Delta\equiv\nabla^2$ is the Laplacian operator. We introduced in Eq.~(\ref{coulomb}) a spatially varying dielectric permittivity $\e(\br)$. The Yukawa potential is obtained by inverting Eq.~(\ref{Yukawa}) in Fourier space, which yields $w(\br)=\ell_ye^{-b|\br|}/|\br|$. The wall potential $V_w(\br)$ in Eq.~(\ref{Hp}) takes into account the interaction between the particles and the boundaries of the system. It can be used either to model a specific particle-wall interaction, or to restrict the position of particles within a particular region. The self-energy of ions that has to be substracted from the total Hamiltonian is given by
\be
E_s=\frac{q_i^2}{2}v_c^b(\br-\br')|_{\br=\br'}+\frac{1}{2}w(\br-\br')|_{\br=\br'},
\ee
where the bare Coulomb potential $v_c^b(\br-\br')$ is the inverse of the Laplace operator
\be\label{coulombB}
{v^b_c}^{-1}(\br,\br')=-\frac{k_BT\e_w}{e^2}\Delta\delta(\br-\br').
\ee
The electrostatic potential, solution of Eq.~(\ref{coulombB}), has the well-known form $v_c^b(\br)=\ell_w/|\br|$, where $\ell_w=e^2/(4\pi\e_w k_BT)\simeq 7$ {\AA} is the Bjerrum length in water solvent at temperature $T=300$ K, $\e_w=78\hspace{0.5mm}\e_0$ stands for the dielectric permittivity of water and $e$ denotes the elementary charge. The partition function of the system can be put in a more tractable form by performing two Hubbard-Stratanovitch transformations in order to linearize the pairwise interaction terms  within the exponential of Eq.~(\ref{PartCan}) at the cost of introducing two fluctuating auxiliary potentials for each interaction, namely an electrostatic potential $\phi$ and a Yukawa potential $\psi$. The grand canonical partition function defined according to $Z_G=\prod_{i=1}^p\sum_{N_i\geq0}e^{\mu_i N_i}Z_c$ can be recast in this way  in the form of a functional integral over these two potentials~\cite{DunYuk},
\be\label{Zg}
Z_G=\int \mathcal{D}\phi\mathcal{D}\psi\;e^{-H[\phi,\psi]}
\ee
with the functional Hamiltonian
\be\label{HamFunc}
H[\phi,\psi]=\int
\mathrm{d}\br\left[\frac{\left[\nabla\phi(\br)\right]^2}{8\pi\ell_B(\br)}-i\sigma_s(\br)\phi(\br)\right]
+\int\frac{\mathrm{d}\br}{8\pi\ell_y}\left[\left[\nabla\psi(\br)\right]^2+b^2\psi^2(\br)\right]
-\sum_i\lambda_i \int\mathrm{d}\br e^{E_s-V_w(\br)+i \left[q_i\phi(\br)+\psi(\br)\right]}.
\ee
In Eq.~(\ref{HamFunc}), we introduced the spatially varying Bjerrum length $\ell_B(\br)=e^2/\left[4\pi\e(\br)k_BT\right]$ and the rescaled particle fugacity $\lambda_i=e^{\mu_i}/\lambda_T^3$.

\section{Bulk ionic liquid}
\label{BulkLiq}

In this section, we analyze the electrolyte model of Eq.~(\ref{HamFunc}) within the variational formalism in a bulk medium. The absence of boundaries in the system means a vanishing wall potential $V_w(\br)=0$. By taking advantage of the spherical symmetry within the homogeneous medium, we first introduce the Fourier-transformed potentials in the form
\bea
&&\phi(\br)=\int\frac{\mathrm{d}^3\bq}{(2\pi)^3}e^{i\bq\cdot\br}\tphi(q),\hspace{2mm}
\psi(\br)=\int\frac{\mathrm{d}^3\bq}{(2\pi)^3}e^{i\bq\cdot\br}\tpsi(q)\nonumber\\
&&v_c(\br-\br')=\int\frac{\mathrm{d}^3\bq}{(2\pi)^3}e^{i\bq\cdot(\br-\br')}\tv_c(q),\hspace{2mm}
w(\br-\br')=\int\frac{\mathrm{d}^3\bq}{(2\pi)^3}e^{i\bq\cdot(\br-\br')}\tw(q)\nonumber
\eea
with $\tv_c=4\pi\ell_w/q^2$ and $\tw=4\pi\ell_y/(q^2+b^2)$. By injecting these expansions into the total Hamiltonian Eq.~(\ref{HamFunc}) with $\ell_B(\br)=\ell_w=const$, one obtains
\be
H=\frac{1}{2}\int\frac{\mathrm{d}^3\bq}{(2\pi)^3}\left[\tphi(q)\tv_c^{-1}(q)\tphi(-q)+\tpsi(q)\tw^{-1}(q)\tpsi(-q)\right]\nonumber\\
-\sum_i\lambda_{b,i}e^{E_S}\int\mathrm{d}\br
e^{i\int\frac{\mathrm{d}^3\bq}{(2\pi)^3}\left[q_i\tphi(q)+\tpsi(q)\right]e^{i\bq\cdot\br}}.
\ee
In order to simplify the field theoretic calculation that will follow, we re-express this Hamiltonian in a matrix form
\be\label{HamOr}
H=\frac{1}{2}\int\frac{\mathrm{d}^3\bq}{(2\pi)^3}\tbpsi^T\tbV^{-1}\tbpsi-\sum_i\lambda_{b,i}e^{E_S}\int\mathrm{d}\br e^{i\int\frac{\mathrm{d}^3\bq}{(2\pi)^3}\tJ_i^T\tbpsi}
\ee
where the $q$-dependence of the functions within the integrals in Fourier space is implicit. We have defined above the following matrix and vectors
\be
\tbV=\left(
  \begin{array}{cc}
    \tv_c(q)  & 0 \\
    0 & \tw(q)  \\
  \end{array}\right),\\
  \hspace{3mm}
  \tbpsi=\left(
  \begin{array}{c}
   \tphi(q)  \\
   \tpsi(q)  \\
     \end{array}
      \right),
  \hspace{3mm}
  \tJ_i=e^{-i\bq\cdot\br}\left(
  \begin{array}{c}
   q_i  \\
   1  \\
     \end{array}
      \right).\nonumber\\
\ee
Because the Hamiltonian functional~(\ref{HamOr}) is non-linear in $\phi$ and $\psi$, an exact evaluation of the partition function is not possible. We will thus find the upper boundary to the exact grand canonical potential by making a variational ansatz. We choose a trial Hamiltonian of the most general quadratic form
\be\label{varhamB}
H_0=\int\frac{\mathrm{d}^3\bq}{(2\pi)^3}\left[\frac{1}{2}\tbpsi^T\tbG^{-1}\tbpsi-i\tJ_v^T\tbpsi\right]
\ee
with the variational kernel
\be\label{TrialKer}
\tbG^{-1}=\left(
  \begin{array}{cc}
    \tG_0^{-1}  & \tG_c^{-1}\\
    \tG_c^{-1} & \tG_y^{-1}  \\
  \end{array}\right).
\ee
The reference Hamiltonian~(\ref{varhamB}) contains a variational external potential vector $\mathbf{J_v}$ that will be defined below as well as the inverse of three trial potentials : the electrostatic potential $G_0(\br-\br')$, the core potential $G_y(\br-\br')$ and finally the operator $G_c(\br-\br')$ that couples the fluctuating electrostatic and Yukawa potentials. The variational grand potential that will be optimized with respect to these trial functions reads $\Omega_1=\Omega_0+\lan H-H_0\ran_0$, where $\Omega_0=-\ln\int \mathcal{D}\phi\mathcal{D}\psi\;e^{-H_0[\phi,\psi]}$ and the bracket $\lan\cdot\ran_0$ means that the statistical average should be taken with the reference Hamiltonian $H_0$. A standard field-theoretic calculation yields
\be\label{BulkGr}
f_1=\frac{\Omega_1}{V}=-\frac{1}{2}\int\frac{\mathrm{d}^3\bq}{(2\pi)^3}F(q)-
\sum_i\frac{\lambda_{b,i}}{V}\int\mathrm{d}\br e^{-\frac{1}{2}\int\frac{\mathrm{d}^3\bq}{(2\pi)^3}H_i(q)}
\ee
where we introduced the following functions
\bea\label{F1}
&&F(q)=\mbox{Tr}\left(\mathbf{I}-\tbG\tbV^{-1}+\ln\tbG\right)+\tJv^T\tbG\tbV^{-1}\tbG\tJv\hspace{1cm}\\
\label{H1}
&&H_i(q)=\tJ_i^T\tbG\tJ_i+\tJv^T\tbG\tJ_i+\tJ_i^T\tbG\tJv.
\eea
$\mathbf{I}$ stands for the $2\times2$ identity matrix and the trace operator acting on a general matrix $\mathbf{M}$ of the same size is defined as $\mbox{Tr}(\mathbf{M})=\sum_{n=1,2}M_{nn}$. Let us note at his stage that the average value of the external fields computed with the Hamiltonian~(\ref{varhamB}) is given by $\tbpsi^T_0=\left(\tphi_b^*,\tpsi_b^*\right)=\lan\tbpsi^T\ran_0=i\tJv^T\tbG$. By injecting the inverse of this relation that gives $\tJv$ in terms of the variational potentials and Eq.~(\ref{TrialKer}) into Eqs.~(\ref{F1}) and~(\ref{H1}), one obtains
\bea
&&F(q)=2-\frac{\tv_c^{-1}\tG_y^{-1}+\tw^{-1}\tG_0^{-1}}
{\tG_0^{-1}\tG_y^{-1}-\tG_c^{-2}}-\ln\left(\tG_0^{-1}\tG_y^{-1}-\tG_c^{-2}\right)
+\frac{1}{V}\left[\tw^{-1}\tpsi_b(q)\tpsi_b(-q)+\tv_c^{-1}\tphi_b(q)\tphi_b(-q)\right]\\
&&H_i(q)=\frac{\tG_0^{-1}+q_i^2\tG_y^{-1}-2q_i\tG_c^{-1}}{\tG_0^{-1}\tG_y^{-1}-\tG_c^{-2}}-\tw-q_i^2\tv_c^b
+\left[\tpsi_b(-q)+q_i\tphi_b(-q)\right]e^{i\bq\cdot\br}+\left[\tpsi_b(q)+q_i\tphi_b(q)\right]e^{-i\bq\cdot\br}.\nonumber\\
\eea
The external electrostatic and Yukawa potentials follow from the variational equations $\delta f_1/\delta\tphi_b(q)=0$ and $\delta f_1/\delta\tpsi_b(q)=0$. By taking into account the electroneutrality condition $\sum_i\rho_{b,i}q_i=0$, these equations yield
\bea\label{ExEl}
&&\phi_b=0\\
\label{ExYuk}
&&\psi_b=\frac{4\pi\ell_y}{b^2}\sum_i\rho_{b,i}
\eea
where
\be\label{denfug}
\rho_{b,i}=-\lambda_{b,i}\frac{\partial f_1}{\partial\lambda_{b,i}}=\lambda_{b,i}e^{-\frac{1}{2}\int\frac{\mathrm{d}^3\bq}{(2\pi)^3}H_i(q)}
\ee
is the average density of the charged Yukawa liquid. The remaining variational equations $\delta f_1/\delta\tG_0^{-1}=0$, $\delta f_1/\delta\tG_y^{-1}=0$ and $\delta f_1/\delta\tG_c^{-1}=0$ whose solution form with Eqs.~(\ref{ExEl}) and~(\ref{ExYuk}) an upper boundary to the exact grand canonical potential now read
\bea
\label{Vareq1}
&&\tG_y^{-2}\left[J+\tv_c^{-1}-\tG_0^{-1}\right]+\tG_c^{-2}\left[\rho_{tot}+\tw^{-1}+\tG_y^{-1}\right]=0\nonumber\\
\label{Vareq2}
&&\tG_0^{-2}\left[\rho_{tot}+\tw^{-1}-\tG_y^{-1}\right]+\tG_c^{-2}\left[J+\tv_c^{-1}+\tG_0^{-1}\right]=0\nonumber\\
\label{Vareq3}
&&\tG_c^{-1}\left\{\tG_c^{-2}+\tG_y^{-1}\left[J+\tv_c^{-1}\right]
+\tG_0^{-1}\left[\rho_{tot}+\tw^{-1}-\tG_y^{-1}\right]\right\}=0\nonumber
\eea
with the total density $\rho_{tot}=\sum_i\rho_{b,i}$ and the ionic strength $J=\sum_i\rho_{b,i}q_i^2$. With a little bit of algebra, one finds that the only solution to the above system of equations is
\bea\label{bulksol}
\label{varsol1}
&&\tG_c^{-1}=0\\
\label{varsol2}
&&\tG_0^{-1}=\frac{q^2+\kappa_{DH}^2}{4\pi\ell_w}\\
\label{varsol3}
&&\tG_y^{-1}=\frac{q^2+\kappa_{yb}^2}{4\pi\ell_y}
\eea
where we have introduced the Debye-H\"uckel (DH) screening parameter $\kappa_{DH}^2=4\pi\ell_w\sum_i\rho_{b,i}q_i^2$ and defined a new screening parameter
\be\label{Kapy}
\kappa_{yb}^2=b^2+4\pi\ell_y\sum_i\rho_{b,i}
\ee
for the Yukawa potential.

We note that by expanding Eq.~(\ref{HamFunc}) at the quadratic order in $\phi$ and $\psi$, one can see that  at the level of the Gaussian theory, the electroneutrality condition $\sum_i\rho_{b,i}q_i=0$ imposes a vanishing coupling between these two potentials. According to Eq.~(\ref{varsol1}), this coupling is also absent at the variational level. Furthermore, Eq.~(\ref{varsol2}) indicates that the electrostatic potential in the bulk system is identical to the DH potential, that is, it is not modified by core interactions. But the HC potential given now by  Eq.~(\ref{varsol3}) gets an additional screening from the surrounding particles. It is important to note that in contrast with the interactions of electrostatic origin which are screened by the charge density, core interactions are screened by the particle density and this screening survives in the case of neutral particles ($q_i=0$). In the next section, we will show that the modification of this screening at the boundaries has important consequences on the distribution of particles close to solid interfaces. Finally, we note that by taking into account the equations~(\ref{varsol1}-\ref{varsol3}), the relation Eq.~(\ref{denfug}) between the particle fugacity and the bulk density becomes
\bea\label{denBfug}
\rho_{b,i}=\lambda_{b,i}e^{\frac{\ell_y}{2}(\kappa_{yb}-b)+\frac{q_i^2}{2}\ell_w\kappa_{DH}-\psi_b}.
\eea

\section{Slit geometries}
\label{SlGe}
\begin{figure}
\includegraphics[width=0.6\linewidth]{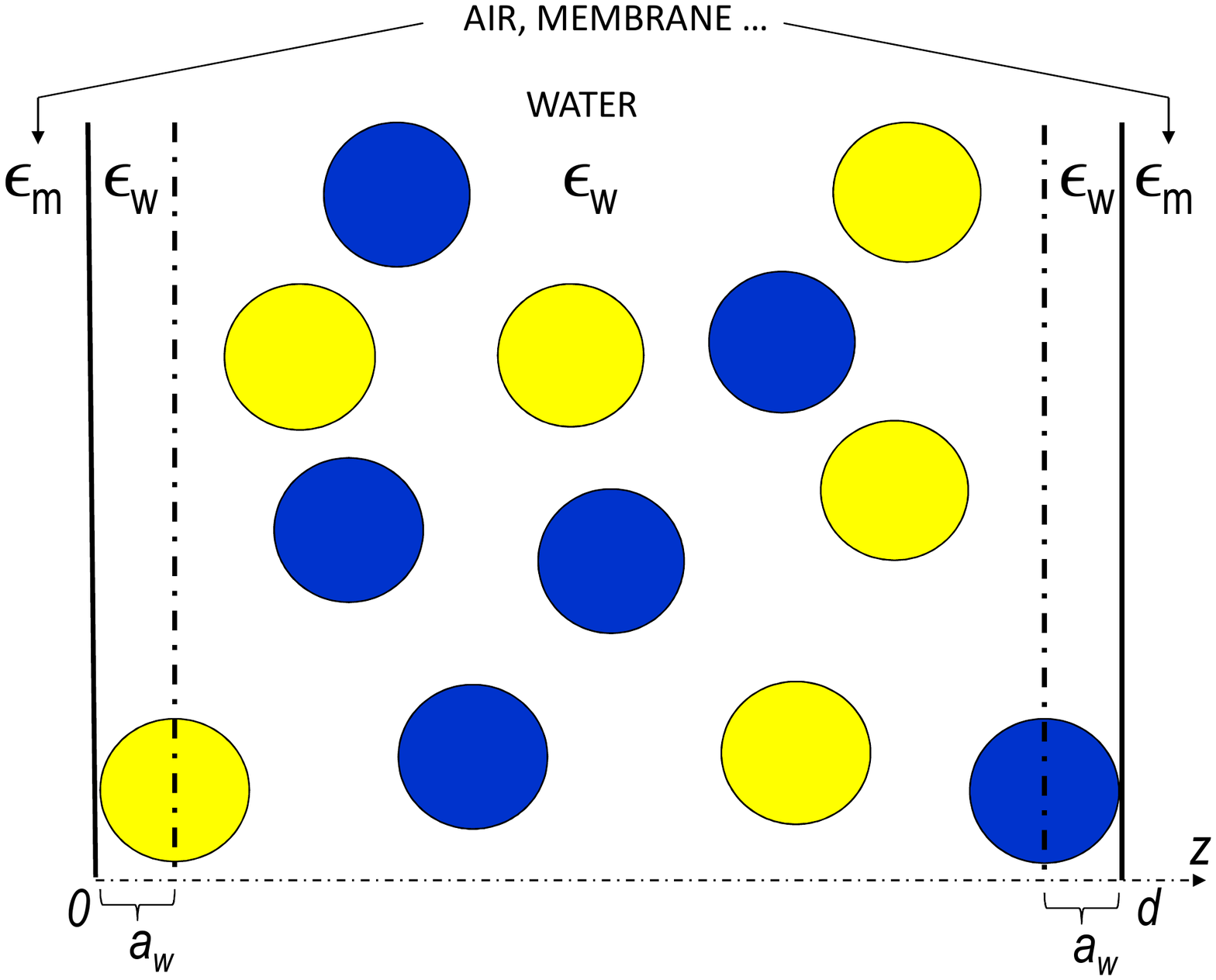}
\caption{(Color online) Geometry for a slit-like pore of thickness $d$. The dielectric permittivities of the pore and the substrate are $\e_w$ and $\e_m$, respectively. Ions occupy the region $a_w<z<d-a_w$, where $a_w$ is the width of the Stern layer.}
\label{sketch}
\end{figure}
In this section, we investigate the effect of finite ion size on the configuration of ions at planar dielectric interfaces as well as on the ionic exclusion from slit pores. The slit pore of thickness $d$ is characterized by a piecewise dielectric permittivity profile
\be\label{PerPr}
\e(z)=\ew\theta(z)\theta(d-z)+\eme[\theta(z)+\theta(z-d)],
\ee
where $\theta(z)$ stands for the Heaviside distribution and $\ew$ and $\eme$ are the dielectric permittivity of water and membrane, respectively (see Fig.~\ref{sketch}). The pore is filled up with charged Yukawa particles whose confinement geometry is imposed with a steric potential defined as
\bea\label{StPot}
&&V_w(z)=0\hspace{0.5mm}\mbox{,}\hspace{4mm}a_w\leq z\leq d-a_w\nonumber\\
&&V_w(z)=\infty\hspace{0.5mm}\mbox{,}\hspace{4mm}z<a_w\hspace{1mm}\mbox{and}\hspace{2mm}z>d-a_w,
\eea
where we introduced the width of the Stern layer $a_w$ (the same for all ionic species) corresponding to a distance of minimum approach to the interface. In this work, we will consider exclusively the case of symmetric electrolytes composed of two species, with $\rho_b^+=\rho_b^-=\rho_b$ and $q_+=-q_-=q$, where + is for cations and - for anions. Furthermore, the pore is in contact with an external particle reservoir at the extremities and the particles within the pore are at chemical equilibrium with the reservoir, which fixes their fugacity according to $\lambda_i=\lambda_{b,i}$. One thus obtains with Eq.~(\ref{denBfug}) the equation that relates the fugacity of the particles within the pore to their reservoir density in the form
\bea\label{denBfugII}
\lambda_i=\rho_{b,i}e^{\psi_b-\frac{\ell_y}{2}(\kappa_{yb}-b)-\frac{q_i^2}{2}\ell_w\kappa_{DH}}.
\eea

The parameters of the Yukawa potential can be indeed determined from neutron-scattering and X-ray experiments that measure the radial distribution function of liquids~\cite{HansBK}. In this article the model parameters will be fixed from simple energetic arguments~\cite{DunYuk}. The dissipation length characterizing the range of the core interactions is usually chosen as a fraction of the ion diameter~\cite{Leb1,Orl1}. We will thus take $b^{-1}=a_i$, where $a_i$ stands for the ion radius. Furthermore, the strength of the Yukawa potential will be determined such that when two Yukawa particles are in contact, the interaction energy is on the order of the thermal energy, i.e. $V_y\left(|\br-\br'|=2a_i\right)=k_BT/2$, which fixes the intensity of the Yukawa potential according to $\ell_y=e^2a_i\simeq7.39\hspace{0.5mm}a_i$. With this requirement, a closer approach of ions will be energetically unfavorable.

\subsection{Calculation schemes}

The field theoretic model Eq.~(\ref{HamFunc}) was studied in Ref.~\cite{DunYuk} for two charged spherical colloids immersed in a counterion liquid. The authors solved the MF equations with a lattice MC procedure. Correlation effects neglected at the MF level can be partially captured with the RPA closure of Ornstein-Zernike (OZ) equations, introduced in Ref.~\cite{Holovko}. Below, we will first explain the derivation of the MF equations and the associated boundary conditions for the general case of charged particles and boundaries carrying a fixed charge distribution. We will then report the prediction from the recently developed RPA theory for the same system~\cite{Holovko}. An alternative calculation method consists in making a variational Ansatz for the reference Hamiltonian and finding the optimal choice for the trial parameters from the minimization procedure. Density profiles obtained within these calculation schemes will be compared with MC simulations in Sec.~\ref{1intNT}  for the case of neutral particles in contact with a planar wall.

\subsubsection{Mean-Field limit and RPA approach}

The mean-field (MF) solution of the grand canonical partition function $Z_G$ associated with the weak surface charge and low density limit is obtained from the saddle point equations $\delta H/\delta\phi(\br)=0$ and $\delta H/\delta\psi(\br)=0$~\cite{DunYuk}, which read
\bea
\label{MFphi}
&&\Delta\phi(\br)-\kappa_{DH}^2e^{-V_w(\br)-\psi(\br)+\psi_b}\sinh\phi(\br)=-4\pi q\ell_w\sigma_s(\br)\\
\label{MFpsi}
&&\Delta\psi(\br)-b^2\psi(\br)+8\pi\ell_y\rho_be^{-V_w(\br)-\psi(\br)+\psi_b}\cosh\phi(\br)=0,
\eea
where $\kappa_{DH}^2=8\pi\ell_wq^2\rho_b$, $\psi_b=8\pi\ell_y/b^2$ and we have also defined $\bar\psi(\br)=-i\psi(\br)$, $\bar\phi(\br)=-iq\phi(\br)$ and dropped the bar over the potentials in order to simplify the notation. In the case of a charged HC liquid confined between two planar surfaces located at $z=0$ and $z=d$, with surface charge $\sigma(z)=\sigma_s\left[\delta(z)+\delta(z-d)\right]$, the differential equations~(\ref{MFphi}) and~(\ref{MFpsi}) should be solved in the region $a_w<z<d-a_w$ with the following boundary conditions (see Appendix~\ref{BounCon})
\bea\label{boundphi2}
&&\left.\frac{d\phi}{dz}\right|_{z=a_w^+}=-4\pi q\ell_w\sigma_s\\
\label{boundpsi2}
&&\left.\frac{d\psi}{dz}\right|_{z=a_w^+}=b\psi\left(z=a_w^+\right)\\
\label{boundphi3}
&&\left.\frac{d\phi}{dz}\right|_{z=\left(d-a_w\right)^-}=4\pi q\ell_w\sigma_s\\
\label{boundpsi3}
&&\left.\frac{d\psi}{dz}\right|_{z=\left(d-a_w\right)^-}=-b\psi\left(z=(d-a_w)^-\right).
\eea
In the case of a system composed of a single interface separating a particle free half space and a solvent medium filled up with ions, the boundary conditions that replace Eq.~(\ref{boundphi3}) and ~(\ref{boundpsi3}) read
\bea\label{boundphi4}
&&\phi(z\to\infty)=0\\
\label{boundpsi4}
&&\psi(z\to\infty)=\psi_b,
\eea
where the constant bulk value of the Yukawa potential is given by Eq.~(\ref{ExYuk}).

In this article, we will treat exclusively the case of neutral interfaces, i.e. $\sigma_s=0$. In this limit, the external electrostatic potential $\phi(z)$ vanishes everywhere. In the dilute limit $8\pi\ell_y\rho_b<b^2$, by expanding at the linear level in $\psi(z)-\psi_b$ the remaining MF equation (\ref{MFpsi}), one can easily find a linear solution for the Yukawa potential $\psi(z)$. The equation to be solved reads
\be
\frac{d^2\psi}{dz^2}-\kappa_{yb}^2\psi=-b^2\psi_b(1+\psi_b).
\ee
The solution to this equation satisfying the boundary conditions~(\ref{boundpsi2}) and~(\ref{boundpsi3}) is given by
\be\label{solMF2in}
\psi(z)=\psi_b-\frac{b\psi_b\cosh\left[\kappa_{yb}\left(d/2-z\right)\right]}
{\kappa_{yb}\sinh\left[\kappa_{yb}(d/2-a_w)\right]+b\cosh\left[\kappa_{yb}(d/2-a_w)\right]}.
\ee
For the single interface system that can be recovered in the limit $d\to\infty$, this linear solution reduces to
\be\label{solMF1in}
\psi(z)=\psi_b\left(1-\frac{b}{b+\kappa_{yb}}e^{-\kappa_{yb}(z-a_w)}\right).
\ee
Finally, the density profiles are obtained by injecting the potentials~(\ref{solMF2in}) or~(\ref{solMF1in}) into the MF relation
\be\label{DenMF}
\rho(z)=\rho_be^{\psi_b-\psi(z)}.
\ee

In a recent work, particle density profiles were derived from the solution of the OZ equation with RPA closure~\cite{Holovko} for the single wall system. The RPA result reads
\be\label{RPAden}
\rho_{RPA}(z)=\rho_b\left[1-V_y(z)-\delta\psi_{RPA}(z)\right]
\ee
where the external potential reads
\be
\delta\psi_{RPA}(z)=\psi_{RPA}(z)-4\pi\ell_y\rho_b/b^2=-4\pi\ell_y\rho_b/b^2e^{-bz}.
\ee
We note that the external potential $\psi_{RPA}(z)$ automatically satisfies the boundary condition Eq.~(\ref{boundpsi2}) on the wall. Although the RPA method partially accounts for correlation effects absent at the MF level, we will show in Sec.~\ref{1intNT} that the MF theory provides a better agreement with MC simulations.

\subsubsection{Variational approach}

An alternative approximative way that allows to go beyond the MF level consists in making a variational ansatz in order to compute an upper boundary to the exact grand potential $\Omega=-k_BT\ln Z_G$. In Sec. III where we treated the case of a bulk electrolyte, it was shown that at the variational level, the electroneutrality condition imposes a vanishing coupling between the electrostatic and  Yukawa potentials. Although this result does not imply a vanishing coupling within the pore, a variational Hamiltonian of the form~(\ref{varhamB}) that couples $\phi(\br)$ and $\psi(\br)$ would lead to a very complicated and untractable set of variational equations for the pore system. We thus opt for a restricted trial Hamiltonian that treats $\phi(\br)$ and $\psi(\br)$ separately, namely $H_0=H_{0\phi}+H_{0\psi}$ with
\bea
&&H_{0\phi}=\frac{1}{2}\int_{\br,\br'}\left[\phi(\br)-i\phi_0(\br)\right]
v^{-1}_0(\br,\br')\left[\phi(\br')-i\phi_0(\br')\right]\nonumber\\
&&H_{0\psi}=\frac{1}{2}\int_{\br,\br'}\left[\psi(\br)-i\psi_0(\br)\right]
w^{-1}_0(\br,\br')\left[\psi(\br')-i\psi_0(\br')\right].\nonumber\\
\label{H0phi}
\eea

The reference Hamiltonian~(\ref{H0phi})  contains four trial functions : two external potentials $\phi_0(\br)$ and $\psi_0(\br)$ whose physical origin will be clarified below, an electrostatic potential $v_0(\br,\br')$ and a core potential $w_0(\br,\br')$. The variational grand potential reads $\Omega_v=\Omega_0+\lan H-H_0\ran_0$, where $\Omega_0=-\ln Z_0=-\ln\int \mathcal{D}\phi\mathcal{D}\psi\;e^{-H_0[\phi,\psi]}$ is the gaussian part of the grand potential that corresponds to the vdW theory~\cite{netzvdw}. Since we did not consider an explicit coupling in the reference Hamiltonian $H_0[\phi,\psi]$, the gaussian contribution can be separated into the Coulomb and Yukawa parts as $\Omega_0=\Omega_{0\phi}+\Omega_{0\psi}$, where
\bea\label{GausCoPhi}
&&\Omega_{0\phi}=-\ln\int \mathcal{D}\phi\;e^{-H_{0\phi}[\phi]}\\
\label{GausCoPsi}
&&\Omega_{0\psi}=-\ln\int \mathcal{D}\psi\;e^{-H_{0\psi}[\psi]}.
\eea
Evaluating the functional integrals in $\Omega_v$, the variational grand potential becomes
\bea\label{grandPot}
\Omega_v&=&\Omega_{0\phi}+\Omega_{0\psi}
+\int\mathrm{d}\br\left(-\frac{\left[\nabla\phi_0(\br)\right]^2}{8\pi\ell_B(\br)}+\sigma_s(\br)\phi_0(\br)\right)
-\int\frac{\mathrm{d}\br}{8\pi\ell_y}\left(\left[\nabla\psi_0(\br)\right]^2+b^2\psi_0^2(\br)\right)\nonumber\\
&&+\frac{k_BT}{2e^2}\int\mathrm{d}\br\mathrm{d}\br'\delta(\br-\br')\nabla_\br\e(\br)\nabla_{\br'}v_0(\br,\br')
+\int\frac{\mathrm{d}\br\mathrm{d}\br'}{8\pi\ell_y}\delta(\br-\br')\left(\nabla_\br\nabla_{\br'}+b^2\right)w_0(\br,\br')
-\sum_i\int\mathrm{d}\br\rho_i(\br)
\eea
where we defined the local density as
\be\label{LocRef1}
\rho_i(\br)=\lambda_ie^{-V_w(\br)-\frac{q_i^2}{2}\left[v_0(\br,\br)-v_c^b(0)\right]-\frac{1}{2}\left[w_0(\br,\br)-w(0)\right]}
e^{-\psi_0(\br)-q_i\phi_0(\br)}.
\ee
The variational equations for the trial potentials $\delta\Omega_v/\delta\phi_0(\br)=0$, $\delta\Omega_v/\delta\psi_0(\br)=0$, $\delta\Omega_v/\delta v_0(\br,\br')=0$ and $\delta\Omega_v/\delta w_0(\br,\br')=0$ read
\bea\label{PBgen}
&&\Delta\phi_0(z)+4\pi\ell_B(\br)\sum_i\rho_i(\br)q_i=-4\pi\ell_B(\br) \sigma_s(\br)\\
\label{YukPBgen}
&&\Delta\psi_0(z)-b^2\psi_0+4\pi\ell_y\sum_i\rho_i(\br)=0\\
\label{DHgen}
&&\left[-\nabla(\e(\br)\nabla)+\e(\br)\kappa_c^2(\br)\right]v_0(\br,\br')=\frac{e^2}{k_BT}\delta(\br-\br')\hspace{6mm}\\
\label{YukDHgen}
&&\left[-\Delta+\kappa_y^2(\br)\right]w_0(\br,\br')=4\pi\ell_y\delta(\br-\br')
\eea
where we introduced the spatially varying screening parameters as
\bea
\label{saltDH}
&&\kappa_c^2(\br)=4\pi\ell_B(\br)\sum_i\rho_i(\br)q_i^2\\
\label{saltYU}
&&\kappa_y^2(\br)=b^2+4\pi\ell_y\sum_i\rho_i(\br).
\eea
The coupled self-consistent equations~(\ref{PBgen})-(\ref{YukDHgen}) take into account correlation effects at a non-linear level. Eqs.~(\ref{PBgen}) and~(\ref{DHgen}) were derived within a variational calculation without Yukawa interactions in Ref.~\cite{netz}. Eq.~(\ref{PBgen}) is an extended PB equation whose solution yields the electrostatic potential in the slit pore, dressed by pore-modified correlation effects. Eq.~(\ref{YukPBgen}) provides the local value of the external Yukawa potential $\psi_0(z)$ that quantifies excluded volume effects in the pore. The third relation Eq.~(\ref{DHgen}) is a modified DH equation and Eq.~(\ref{YukDHgen}) takes into account the modification of Yukawa interactions in the slit pore. These type of non-local closure relations are known to be unsolvable even in the restricted case of ions without core interactions~\cite{netz,PRE}. Hence, to make further progress we restrict the form of the pairwise potentials $v_0(\br,\br')$ and $w_0(\br,\br')$ as follows.

Following our work on simple ions~\cite{PRE}, we choose the variational electrostatic potential as the inverse of the generalized DH operator
\be\label{DH1}
v_0^{-1}(\br,\br')=\frac{k_BT}{e^2}\left[-\nabla(\e(\br)\nabla)+\e(\br)\kappa_c^2(\br)\right]\delta(\br-\br')
\ee
where $\e(\br)$ is given by Eq.~(\ref{PerPr}) and $\kappa_c(\br)=\kappa_c\theta(z-a_w)\theta(d-a_w-z)$ is a constant piecewise trial screening parameter which is uniform within the pore. This choice is indeed a generalized Onsager-Samaras approximation~\cite{onsager}. It was shown in our previous work on purely Coulombic liquids that the approximation yields a good agreement with MC simulations for ion distributions in charged pores~\cite{PRE}.

For the sake of physical consistency, the choice of the trial Yukawa potential should be done with care.  In section III, it was shown that Yukawa interactions get an additional screening due to the presence of surrounding particles. Hence, we expect the inverse screening length to reduce to the bare value $b$ in the particle-free regions. This point is also confirmed by the form of the screening parameter Eq.~(\ref{saltYU}) associated with the general variational equations~(\ref{PBgen})-(\ref{YukDHgen}). We thus chose the variational Yukawa potential as the inverse of the following operator
\be\label{DH2}
w_0^{-1}(\br,\br')=\frac{-\Delta+\kappa_y^2(\br)}{4\pi\ell_y}\delta(\br-\br')
\ee
where $\kappa_y(\br)=b[\theta(a_w-z)+\theta(z+a_w-d)]+\kappa_y\theta(z-a_w)\theta(d-a_w-z)$ is an effective screening parameter characterizing the pore-modified screening of core interactions that reduces to the bare screening parameter in the membrane.

The effective electrostatic and Yukawa potentials obtained by inverting Eqs.~(\ref{DH1}) and (\ref{DH2}) in the slit geometry are reported in Appendix~\ref{appendixVarFr}. We are thus left with two variational parameters $\kappa_c$, $\kappa_y$ and two variational potentials $\phi_0(\br)$ and $\psi_0(\br)$. The corresponding variational grand potential reads
\bea\label{grandPot}
\Omega_v&=&\Omega_{0\phi}+\Omega_{0\psi}
+S\int\mathrm{d}z\left(-\frac{\left[\nabla\phi_0(z)\right]^2}{8\pi\ell_B(z)}+\sigma_s(z)\phi_0(z)\right)
-S\int\frac{\mathrm{d}z}{8\pi\ell_y}\left(\left[\nabla\psi_0(z)\right]^2+b^2\psi_0^2(z)\right)\\
&&-S\int_{a_w}^{d-a_w}\mathrm{d}z\left[\frac{\kappa_c^2}{8\pi\ell_w}v_0(\br,\br)
+\frac{\kappa_y^2-b^2}{8\pi\ell_y}w_0(\br,\br)\right]
-S\sum_i\int_{a_w}^{d-a_w}\mathrm{d}z\rho_i(z)
\eea
where $S$ stands for the lateral surface. For charged Yukawa particles confined in a slit pore, by taking into account the chemical equilibrium condition Eq.~(\ref{denBfug}) and defining the following potentials
\bea\label{PotEl}
&&V_c(z)=\frac{q^2}{2}\left[\ell_w(\kappa_{DH}-\kappa_c)+\delta v_0(z)\right]\\
\label{PotYu}
&&V_y(z)=\frac{1}{2}\left[\ell_y(\kappa_{yb}-\kappa_y)+\delta w_0(z)\right],
\eea
the local particle density Eq.~(\ref{LocRef1}) becomes
\be\label{locden}
\rho_\pm(z)=\rho_be^{-V_w(z)-V_c(z)-V_y(z)+\psi_b-\psi_0(z)\mp\phi_0(z)}.
\ee
For a symmetric electrolyte, by rescaling the electrostatic potential according to $\bar\phi(z)\equiv q\phi(z)$, the variational equations for the external potentials, $\delta\Omega_v/\delta\psi_0(\br)=0$ and $\delta\Omega_v/\delta\bar\phi_0(\br)=0$ read
\begin{widetext}
\bea\label{eqvarEX1}
&&\Delta\phi_0(z)-\kappa^2_{DH}e^{-V_w(z)-V_c(z)-V_y(z)+\psi_b-\psi_0(z)}\sinh\phi_0(z)=-4\pi\ell_w q\sigma_s(z)\\
\label{eqvarEX2}
&&\Delta\psi_0(z)-b^2\psi_0+8\pi\ell_y\rho_be^{-V_w(z)-V_c(z)-V_y(z))+\psi_b-\psi_0(z)}\cosh\phi_0(z)=0
\eea
where we dropped the bar over the electrostatic potential in order to simplify the notation. The differential equations~(\ref{eqvarEX1}) and~(\ref{eqvarEX2}) should be solved with the boundary conditions~(\ref{boundphi2})-(\ref{boundpsi4}). The variational equations for the remaining trial parameters can be obtained in a similar way from the simple derivatives $\partial\Omega_v/\partial\kappa_c=0$ and $\partial\Omega_v/\partial\kappa_y=0$ in the form
\bea\label{eqvarSC1}
&&\kappa_c^2=\kappa^2_{DH}\lan e^{-V_c(z)-V_y(z)+\psi_b-\psi_0(z)}\cosh\phi_0(z)
\frac{\partial V_c}{\partial\kappa_c}\ran_p\lan\frac{\partial V_c}{\partial\kappa_c}\ran_p^{-1}\\
\label{eqvarSC2}
&&\kappa_y^2=b^2+8\pi\ell_y\rho_b\lan e^{-V_c(z)-V_y(z)+\psi_b-\psi_0(z)}\cosh\phi_0(z)
\frac{\partial V_y}{\partial\kappa_y}\ran_p\lan\frac{\partial V_y}{\partial\kappa_y}\ran_p^{-1}
\eea
\end{widetext}
where we have introduced the pore average defined as $\lan\cdot\ran_p=\int_{a_w}^{d-a_w}\mathrm{d}z\cdot/(d-2a_w)$.

The set of coupled variational equations~(\ref{eqvarEX1})-(\ref{eqvarSC2}) is the main theoretical result of this article. The relation Eq.~(\ref{eqvarSC1}) is a modified DH equation for the effective pore screening length. Eq.~(\ref{eqvarSC2}) takes into account the modification of the screening of Yukawa interactions in the slit pore. It is important to note at this stage that these relations are valid in all geometries (spherical, cylindrical ...) and can be solved with an iterative algorithm. We emphasize that Eqs.~(\ref{eqvarEX1}) and~(\ref{eqvarSC1}) were derived for ionic liquids without Yukawa interactions ($\psi(z)=0$ and $V_y(z)=0$) and solved in slit and cylindrical pores in Refs.~\cite{PRE,PRL,JCP}. It should be also noted that approximate self-consistent equations, similar in form to  Eqs.~(\ref{eqvarEX1}) and~(\ref{eqvarSC1}) have been frequently used in nanofiltration studies~\cite{yarosch,Szymczyk,YarII}. The prediction of these approximate equations for salt rejection from neutral and charged pores was compared with that of the variational equations in Ref.~\cite{PRE,JCP} and it was shown that the former equations can overestimate ionic concentrations in cylindrical pores by a factor of 2 to 3. The relations Eqs.~(\ref{eqvarEX1})-(\ref{eqvarSC2}) now generalize these self-consistent equations to the case of electrolytes composed of large ions associated with important excluded volume effects.

The potentials~(\ref{PotEl}) and~(\ref{PotYu}) are obtained in the limit $\br'\to\br$ from the corresponding kernels given by Eqs.~(\ref{kernel1}) and~(\ref{kernel4}). One finds for the slit pore geometry
\bea\label{PMFslit1}
V_c(z)&=&\frac{q^2\ell_w}{2}(\kappa_{DH}-\kappa_c)
+\frac{q^2\ell_w}{2}\int\frac{\mathrm{d}kk}{\rho_c}\frac{\Delta_c}{1-\Delta_c^2e^{-2\rho_c(d-2a_w)}}\left[e^{-2\rho_c(z-a_w)}
+e^{-2\rho_c(d-a_w-z)}+2\Delta_c e^{-2\rho_c(d-2a_w)}\right]\nonumber\\
&&\\
\label{PMFslit2}
V_y(z)&=&\frac{\ell_y}{2}(\kappa_{yb}-\kappa_y)
+\frac{\ell_y}{2}\int\frac{\mathrm{d}kk}{\rho_y}\frac{\Delta_y}{1-\Delta_y^2e^{-2\rho_y(d-2a_w)}}
\left[e^{-2\rho_y(z-a_w)}
+e^{-2\rho_y(d-a_w-z)}+2\Delta_y e^{-2\rho_y(d-2a_w)}\right].
\eea
The functions $\rho_c,\rho_y,\eta_c,\Delta_c,\Delta_y$ are reported in Appendix~\ref{appendixVarFr}. We also introduce the partition coefficient, i.e. the pore-averaged particle density as $k=\lan\rho(z)/\rho_b\ran_p$.

In the case of a simple dielectric interface located at $z=0$, the minimization of $\Omega_v$  at constant fugacity with respect to the inverse screening lengths yields $\kappa_c=\kappa_{DH}$ and $\kappa_y=\kappa_{yb}$. We are thus left with two variational equations~(\ref{eqvarEX1}) and~(\ref{eqvarEX2}) to be solved. For the single interface geometry ($d\to\infty$), the Yukawa potential~(\ref{PMFslit2}) becomes
\be\label{PMF1INT1}
V_y(z)=\frac{\ell_y}{2}\int\frac{\mathrm{d}kk}{\rho_y}\Delta_ye^{-2\rho_y(z-a_w)}.
\ee
and the electrostatic potential~(\ref{PMFslit1}) that we separate into a solvation and a dielectric part, $V_c(z)=V_{cs}(z)+V_{ci}(z)$ reads
\bea\label{PMF1INT2}
&&V_{cs}(z)=\frac{q^2\ell_w}{2}\int\frac{\mathrm{d}kk}{\rho_c}\frac{\rho_c-k}{\rho_c+k}e^{-2\rho_c(z-a_w)}\\
\label{PMF1INT3}
&&V_{ci}(z)=q^2\ell_w\int\frac{\mathrm{d}kk}{\rho_c}\frac{(1-\eta_c)k\rho_c}{(\rho_c+k)(\rho_c+\eta_c k)}e^{-2\rho_c(z-a_w)}.
\eea
We note that in the case of neutral interfaces ($\sigma_s=0$), the external electrostatic potential vanishes, i.e. $\phi_0(z)=0$.

We also computed the surface tension for the charged Yukawa fluid. The surface tension, denoted by $\sigma_e$, is defined as the excess Grand potential, i.e. the difference between the Grand potential of the interfacial system and that of the bulk system. $\sigma_e$ was computed in Ref.~\cite{David} within a first order perturbation theory. The model included a Stern layer induced by the finite ion size. A sophisticated variational calculation of $\sigma_e$ that accounts for a depletion layer induced by electrostatic image forces was also proposed in Ref.~\cite{hatlo}. The derivation of $\sigma_e$ from the variational Grand potential Eq.~(\ref{grandPot}) is explained in Appendix~\ref{SurTenTen}. The result reads
\bea\label{surten}
\beta\sigma_e&=&2\rho_ba_w-\frac{a_w}{24\pi}\left[(\kappa_{yb}-b)(\kappa_{yb}^2+\kappa_{yb}b-2b^2)+\kappa_{DH}^3\right]+\frac{b^2a_w\psi_b^2}{8\pi\ell_y}
+\int\frac{\mathrm{d}kk}{8\pi}\left[\ln\frac{(\rho_c+k\eta_c)^2}{k\rho_c(\eta_c+1)^2}-\frac{\kappa_{DH}^2\Delta_c}{2\rho_c^2}\right]\nonumber\\
&&-\frac{b^2}{16\pi}\ln\frac{(\kappa_{yb}+b)^2}{4b\kappa_{yb}}-\frac{b\psi_0^2(a_w^+)}{8\pi\ell_y}-
\int_{a_w}^\infty\frac{\mathrm{d}z}{8\pi\ell_y}\left[(\nabla\psi_0)^2+b^2(\psi_0^2-\psi_b^2)\right]
-2\int_{a_w}^\infty\mathrm{d}z\left[\rho(z)-\rho_b\right]
\eea
where $\beta=1/k_BT$. The relation~(\ref{surten}) generalizes the result of Ref.~\cite{David} to the case where core collisions are present. The first three terms proportional to $a_w$ include the contribution of the exclusion effect induced by the Stern layer. The next four terms contain corrections from the first order variational calculation to the quadratic fluctuations of the electrostatic and Yukawa potentials. Finally, the last term brings non-linear contributions from the surface depletion. We finally note that the evaluation of Eq.~(\ref{surten}) requires the numerical integration of Eq.~(\ref{eqvarEX2}).

\subsection{Single interface}
\label{1intNT}
\begin{figure}[t]
(a)\includegraphics[width=0.6\linewidth]{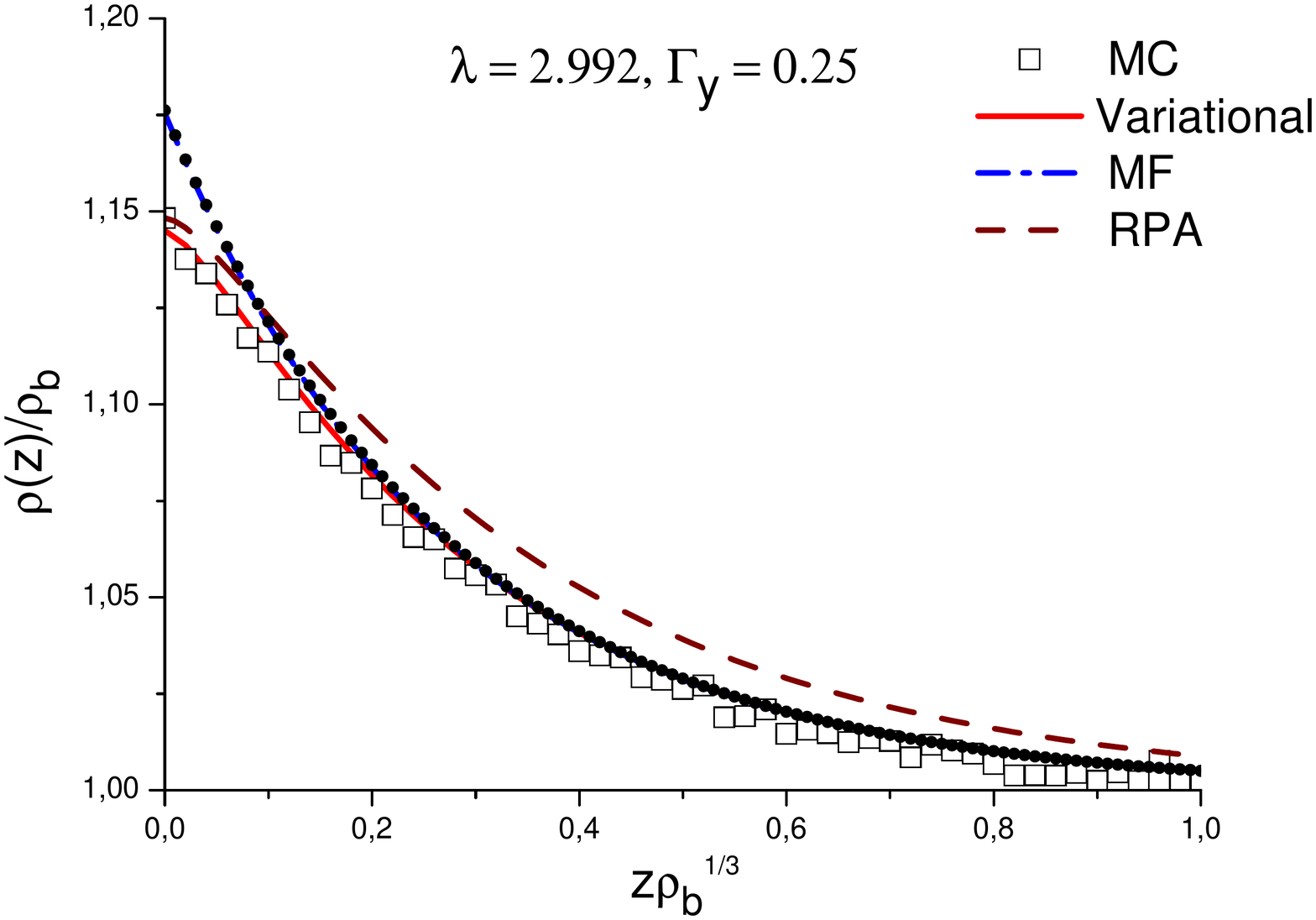}
(b)\includegraphics[width=0.6\linewidth]{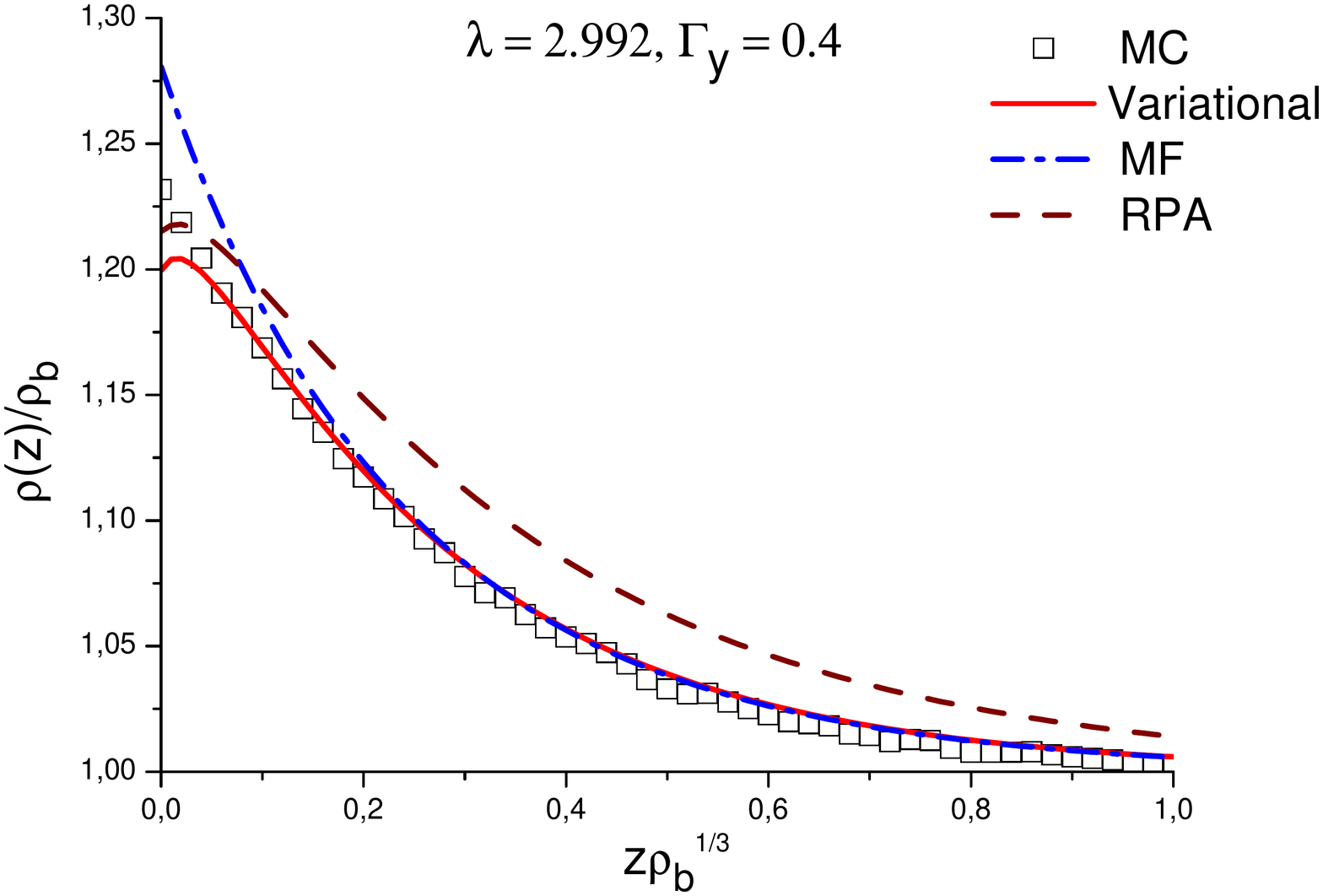}
(c)\includegraphics[width=0.6\linewidth]{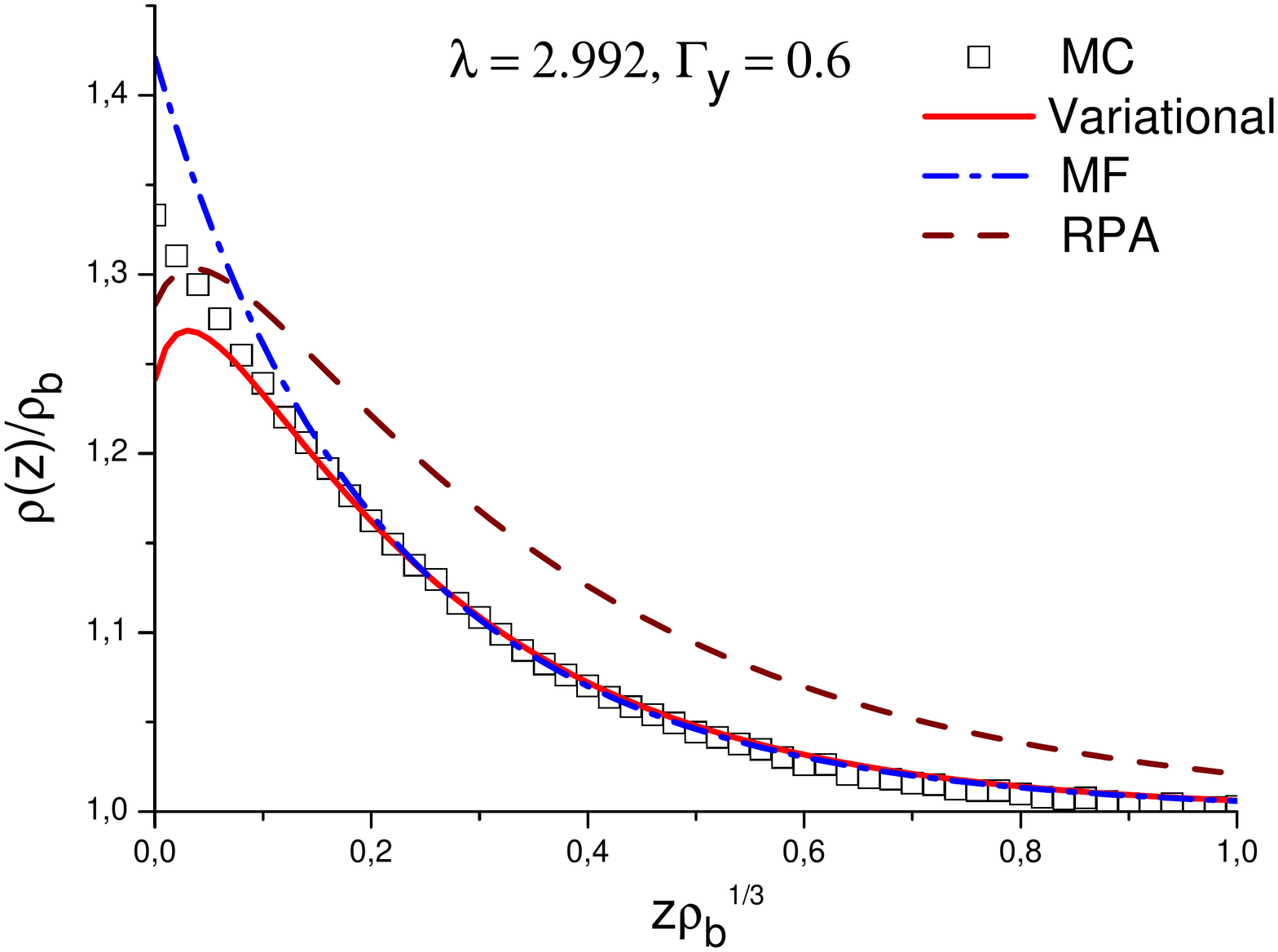}
\caption{(Color online) Density profile of neutral Yukawa particles at the interface for $\lambda_y=2.992$ and (a) $\Gamma_y=0.25$, (b) $\Gamma_y=0.4$ and (c) $\Gamma_y=0.6$. The plots compare MC results (squares) with the variational method (continuous lines), the MF theory (dash-dotted line) and the RPA method (dashed lines). The black circles in (a) are obtained with the linear solution Eq.~(\ref{solMF1in}).}
\label{figMF}
\end{figure}

In this section, we investigate the partition of neutral and charged Yukawa particles in contact with a neutral planar interface located at $z=0$. The single interface system is recovered from the slit geometry depicted in Fig.~\ref{sketch} in the limit $d\to\infty$. The right half space is filled up with ions in water medium of dielectric permittivity $\e_w=78$. The left half space is free of ions and composed of biological matter or air, associated with a low dielectric permittivity $\e_m<\ew$. We will take $\e_m=2$ (the case for lipid membranes), unless otherwise stated. This choice is also a good approximation for the water-air interface characterized by $\e_m=1$. We will first investigate the case with a vanishing Stern layer $a_w=0$ and then discuss the modifications induced by a finite $a_w$.

Fig.~\ref{figMF} displays for $a_w=0$ the distribution of neutral Yukawa particles ($q=0$) for model parameters $\lambda_y\equiv b\rho_b^{-1/3}=2.992$ and $\Gamma_y\equiv\ell_y\rho_b^{1/3}=0.25$, 0.4, and 0.6. We present a comparison of the MF theory, the variational formalism and the RPA method with the result of canonical MC simulations. First of all, one sees that MC data show wetting of the interface. The adsorption of repulsive particles onto a hard wall was also observed in other MC simulations~\cite{Feng,Zhou,HolovkoII} as well as within various theoretical approaches~\cite{Lebowitz1,Weeks,Holovko}. The underlying mechanism is known to be the pressure applied by the bulk on the particles close to the surface, which originates from core collisions. At the MF and variational levels, this effect is incorporated in the attractive potential $\psi(z)-\psi_b$ obtained from the numerical solution of Eqs.~(\ref{MFpsi}) and~(\ref{eqvarEX2}). By comparing the plots of Fig.~\ref{figMF} from top to bottom, one sees that an increase of the coupling of core interactions that amplifies excluded volume effects results in a stronger particle adsorption on the wall. A comparison of the MF level density profile with the simulation data in Fig.~\ref{figMF}  shows that although there exists a good agreement far from the interface, the MF theory slightly overestimates the particle density close to the surface and the discrepancy  grows with increasing $\Gamma_y$. We also note in passing that the analytical solution given in Eq.~(\ref{solMF1in}) of the linearized MF equation in Fig.~\ref{figMF}.a shows a very good agreement with the numerical solution of Eq.~(\ref{MFpsi}).

The overestimation of the particle density at the MF level can be explained by the incapacity of the MF theory to take into account the correlations associated with core interactions that are partially embodied at the variational level in the potential $V_y(z)$ of Eq.~(\ref{PMF1INT1}). In Sec. III, it was shown that in a homogeneous medium, a Yukawa particle surrounded by other particles experiences core interactions reduced by the screening effect. The repulsive potential $V_y(z)$ that reduces the contact density is induced by the modification of this screening close to the boundary : because the left side of the interface is free of particles, the Yukawa particle is more efficiently screened in the bulk than in the neighborhood of the surface. Since the screening is favorable to the system, the particle exhibits a tendency to move away from the interface. As can be seen in Fig.~\ref{figMF}, the variational formalism that can account for this weak solvation effect shows a better agreement with MC results. One notes in Fig.~\ref{figMF}.b and c that very close to the interface, the variational result deviates from the simulation data and exhibits a weak concentration peak absent in MC simulations. This might be due to the simple choice of a uniform $\kappa_y$ in Eq.~(\ref{DH2}). We emphasize that for a bulk concentration $\rho_b=1$ M, the region where the disagreement takes place corresponds to the interval $z\lesssim 0.5$ {\AA}. Finally, we note that the RPA method exhibits a poor agreement with MC simulations over the whole interfacial area and the discrepancy increases with $\Gamma_y$.

\begin{figure}[t]
(a)\includegraphics[width=0.6\linewidth]{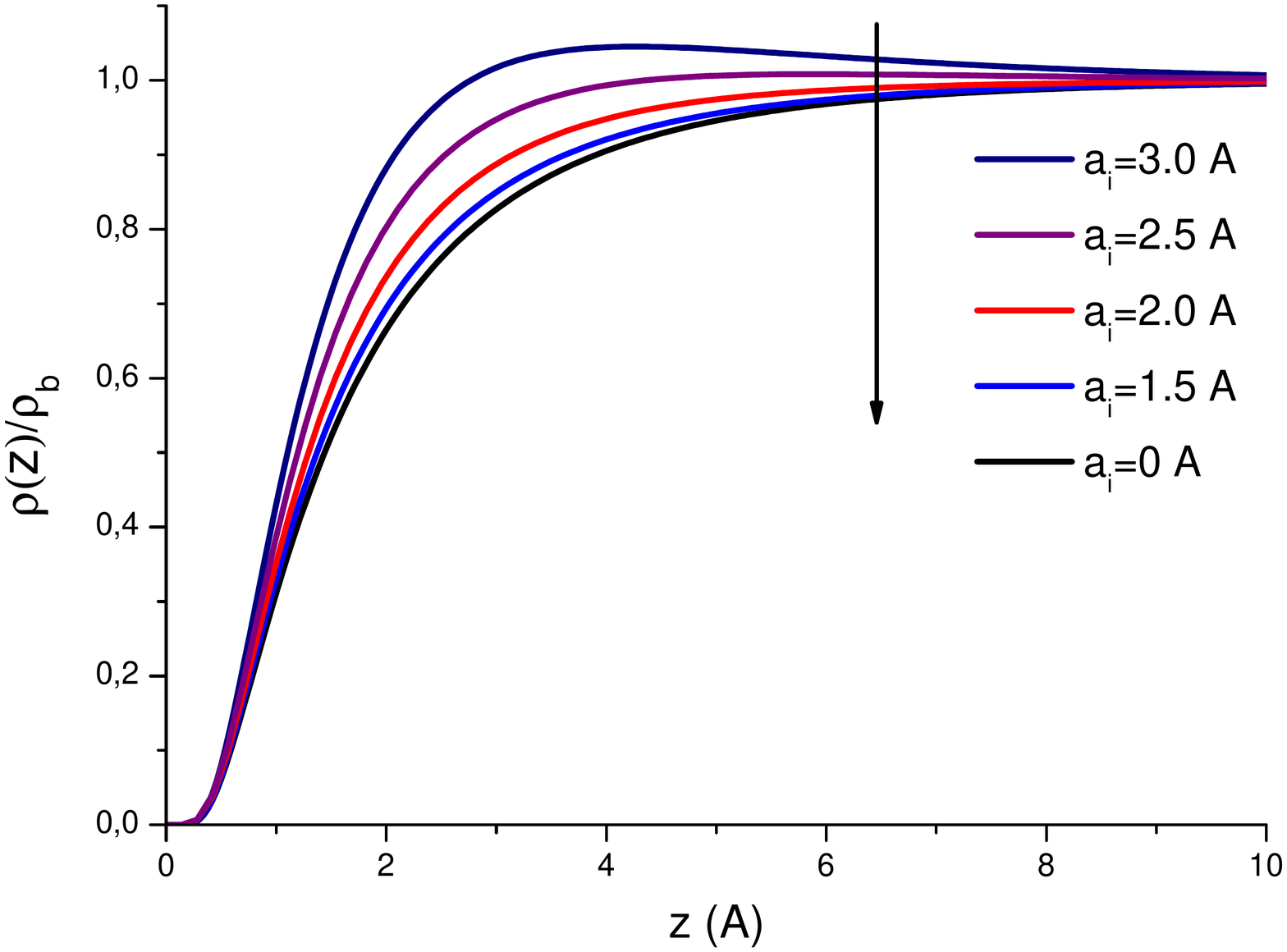}
(b)\includegraphics[width=0.6\linewidth]{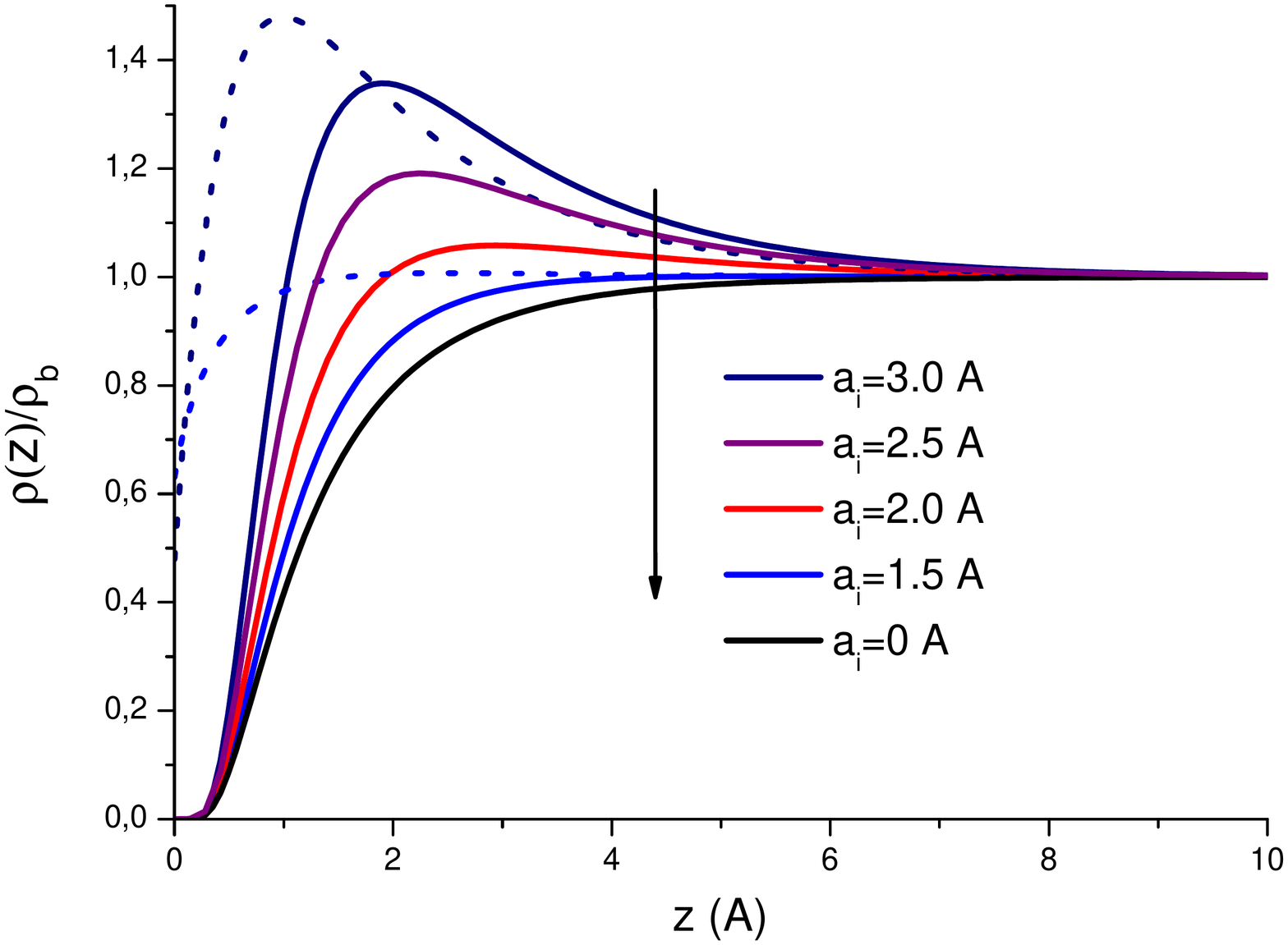}
\caption{(Color online) Density profile of monovalent ions ($q=1$) at the dielectric interface for $a_w=0$, $\e_w=78$, $\e_m=2$ and ion sizes $a_i=0,1.5,2,2.5,3$ {\AA}. Bulk concentrations are (a) $\rho_b=0.3$ M and (b) $\rho_b=1$ M. The dashed curves in the bottom plot shows the density profiles for $a_i=1.5$ {\AA} and $a_i=3.0$ {\AA} for the case $\e_m=\e_w=78$.}
\label{figCHvsn}
\end{figure}
\begin{figure}[t]
\includegraphics[width=0.6\linewidth]{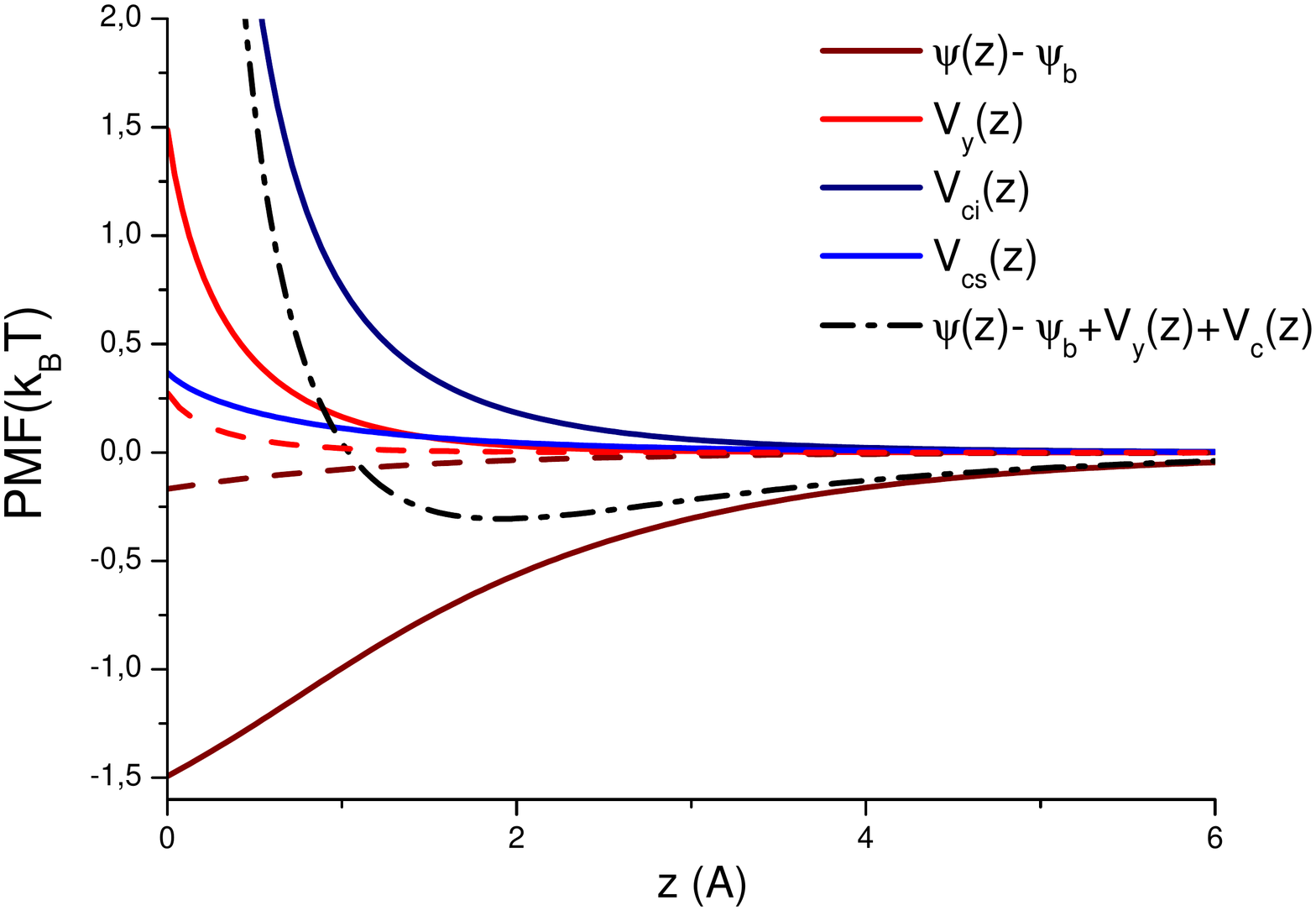}
\caption{(Color online) Solid lines illustrate for $\rho_b=1$ M, $a_i=3.0$ {\AA} and $a_w=0$ the behaviour of the potentials $\psi(z)-\psi_b$, $V_y(z)$, $V_{cs}(z)$, $V_{ci}(z)$ and the total PMF in units of $k_BT$. Brown and red dashed lines show $\psi(z)-\psi_b$ and $V_y(z)$, respectively,  for a smaller ion size of $a_i=1.5$ {\AA}.}
\label{PMF}
\end{figure}
We illustrate in Fig.~\ref{figCHvsn} the density of monovalent ions ($q=1$) in the neighborhood of a dielectric interface (e.g. air-water or protein-water interface) for two bulk concentrations and four ion sizes. We note that in the case of particles with a finite charge, Eq.~(\ref{eqvarEX2}) is integrated by including the electrostatic potential $V_c(z)$  given by Eq.~(\ref{PMF1INT2}) and~(\ref{PMF1INT3}). This potential contains image-charge interactions induced by the dielectric discontinuity through the interface and an electrostatic solvation effect. The latter originates from the absence of ions in the region $z<0$ that modifies the electrostatic screening close to the boundary. Both effects are known to be repulsive and in the case of point like ions without core interactions, they lead to an ionic exclusion layer at the interface and a density profile that monotonically increases towards $\rho_b$ with increasing distance from the surface~\cite{hatlo,PRE}. This behaviour is illustrated in Fig.~\ref{figCHvsn} by solid black curves ($a_i=0$). It is well established that the same image forces contain the leading contribution to the positive surface tension of electrolytes at the water-air interface~\cite{onsager} and the exclusion of ions from confined pores~\cite{yarosch,hatlo,PRE,PRL}. By comparing the density profile of point-like ($a_i=0$) and finite size ions ($a_i>0$), one notices that core collisions lead to a net increase of the ion density in the interfacial region. For ions of size $a_i\lesssim 2$ {\AA}, the density profile $\rho(z)$ keeps its monotonic trend whereas for larger ions ($a_i\gtrsim 2$ {\AA}), $\rho(z)$ exhibits an oscillatory shape. With increasing distance from the surface, $\rho(z)$ exceeds the bulk density and exhibits a maximum at a characteristic distance. Beyond this distance, the ion concentration begins to decrease towards the bulk limit. Moreover, we notice in Fig.~\ref{figCHvsn} that ionic adsorption becomes more pronounced with increasing bulk density. Specifically, for a bulk concentration $\rho_b=1$ M, the amplification of the concentration peak that accompanies the increase of the ion size from $a_i=0$ {\AA} to $a_i=3.0$ {\AA} is four times higher than in the more dilute case $\rho_b=0.3$ M.
\begin{figure}[t]
(a)\includegraphics[width=0.6\linewidth]{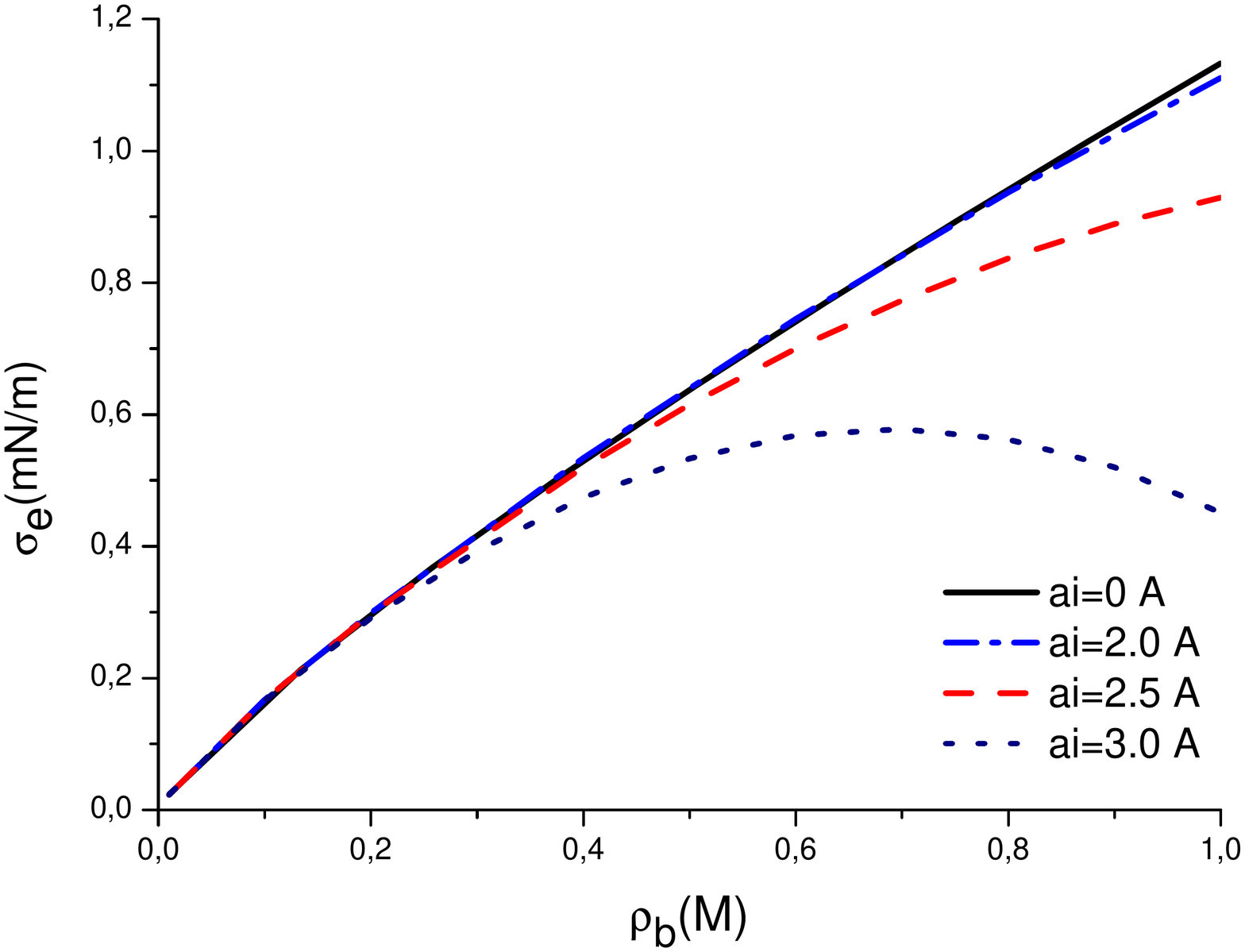}
(b)\includegraphics[width=0.6\linewidth]{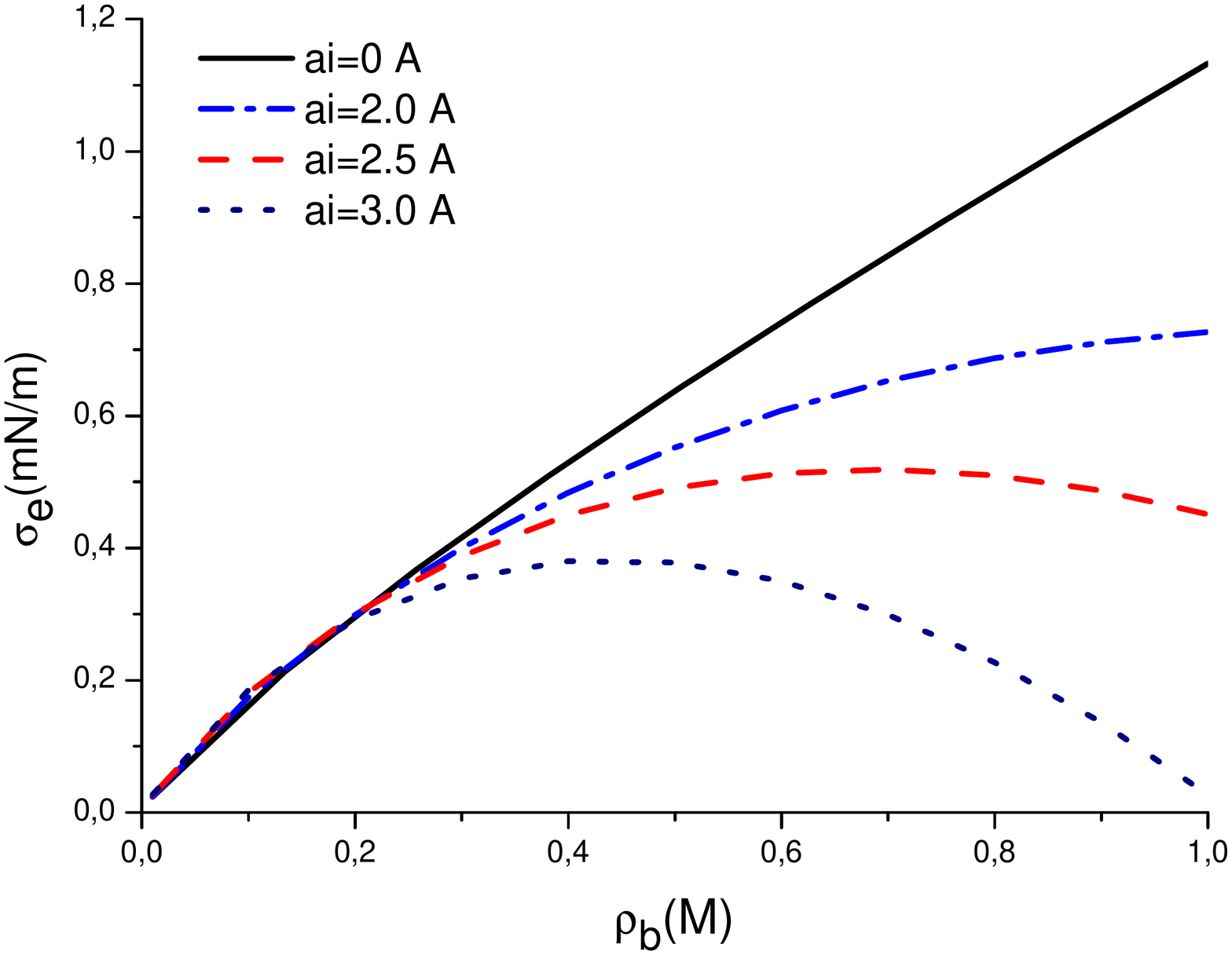}
\caption{(Color online) Surface tension of monovalent ions against the bulk concentration for various ion sizes, $\e_m=2$, (a) $a_w=0$ and (b) $a_w=a_i$.}
\label{SurTen1}
\end{figure}

As illustrated in Fig.~\ref{PMF} that shows the behaviour of the electrostatic and Yukawa potentials as well as the evolution of the total potential of mean force (PMF), the non-monotonic shape of the ion density originates from a competition between the combined image-charge and solvation forces (included in the potentials $V_c(y)$ and $V_y(z)$) that drive the charges to the bulk and the positive pressure applied by the bulk particles on the surface particles (the external Yukawa field $\psi_0(z)$) that move them towards the surface. A comparison of the electrostatic and Yukawa potentials for the cases $a_i=1.5$ {\AA} and $a_i=3.0$ {\AA} shows that the intensification of the ionic adsorption with increasing ion size is essentially due to a rise in the amplitude of the attractive potential $\psi(z)-\psi_b$, which in turn originates from an amplification of excluded-volume effects. Although the repulsive potential $V_y(z)$ increases with increasing $a_i$ almost by the same amount as $\psi(z)-\psi_b$ on the surface, we notice that it experiences a stronger screening than the latter. Hence, the effect of the potential $V_y(z)$ comes into play over a short distance. Furthermore, by comparing the amplitude of the potentials in Fig.~\ref{PMF}, one notices that in the limit $a_w=0$ that we have considered so far, the ionic depletion is mainly due to the interaction of ions with their images, i.e. $V_{c,i}(z)$. This is also illustrated in Fig.~\ref{figCHvsn}.(b) where we plot for two particle sizes the density profiles for the case $\e_m=\e_w=78$, where image interactions vanish (dashed lines). It is seen that the depletion layer is greatly reduced and the height of the concentration peak for the ion size $a_i=3$ {\AA} is amplified.
\begin{figure}[t]
(a)\includegraphics[width=0.6\linewidth]{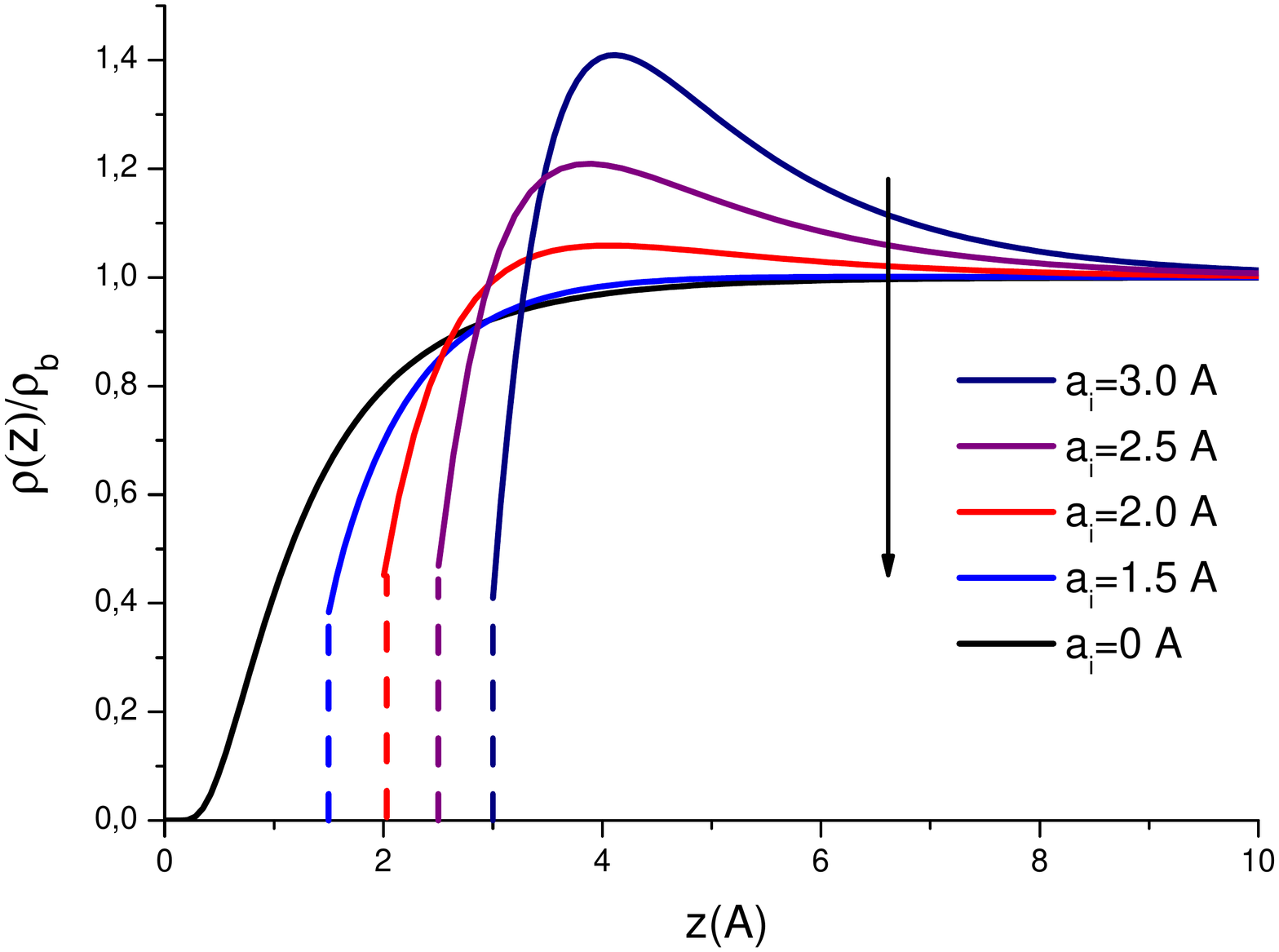}
(b)\includegraphics[width=0.6\linewidth]{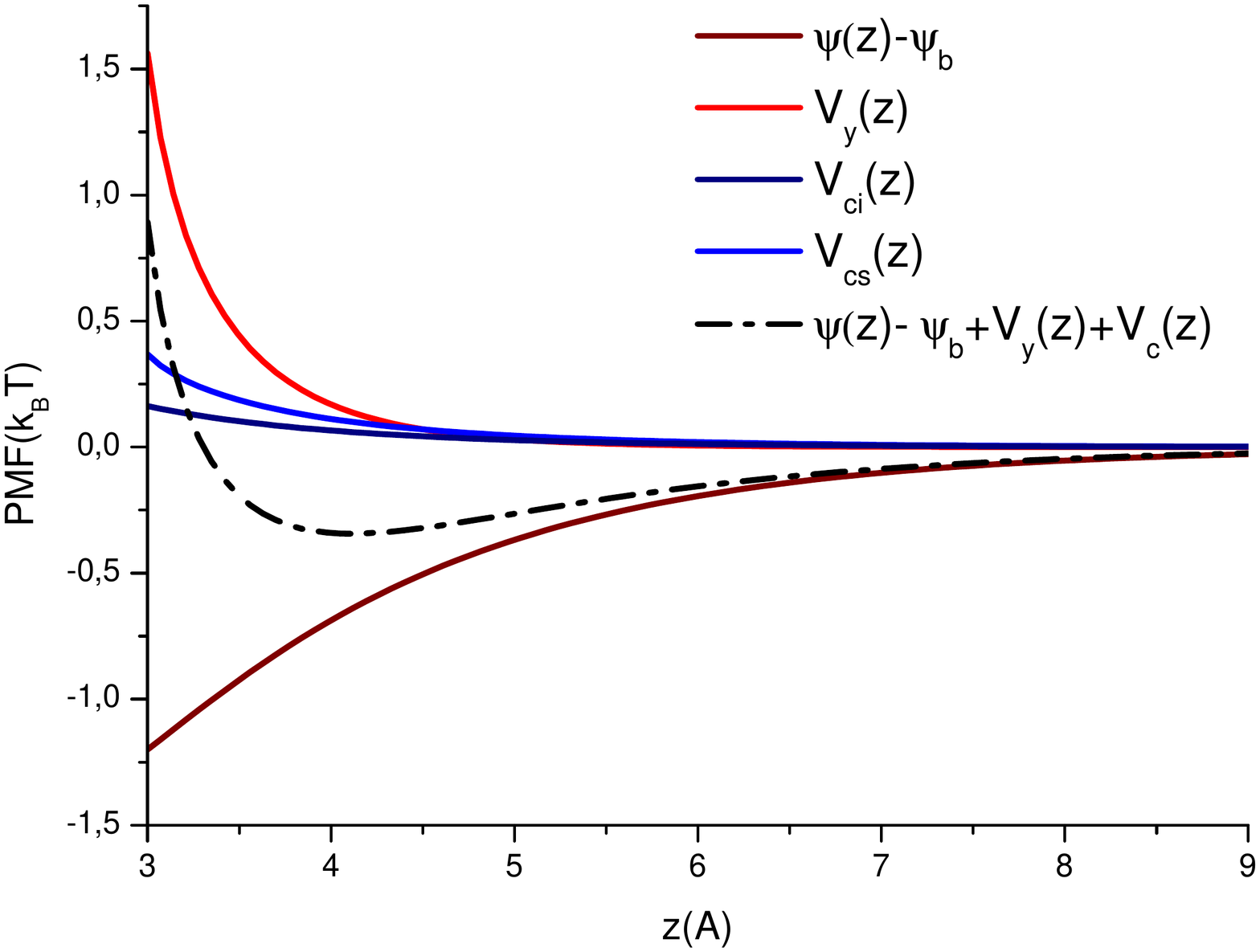}
\caption{(Color online) (a) Density profiles of monovalent ions at the dielectric interface for $a_w=a_i$, $\e_w=78$, $\e_m=2$ and ion sizes $a_i=0,1.5,2.0,2.5,3.0$ {\AA}. (b) The potentials $\psi(z)-\psi_b$, $V_y(z)$, $V_{cs}(z)$, $V_{ci}(z)$ and the total PMF in units of $k_BT$ for $a_w=a_i$ and ion size $a_i=3.0$ {\AA}.}
\label{PMF2}
\end{figure}

We illustrate in Fig.~\ref{SurTen1}.a the evolution of the surface tension $\sigma_e$ Eq.~(\ref{surten}) as a function of the bulk concentration for various ion sizes and the model parameters $\e_m=2$  and $a_w=0$. According to the isothermal Gibbs adsorption equation $\mathrm{d}\sigma_e = -\gamma\mathrm{d}\mu$, where $\mu$ is the chemical potential of the particle and $\gamma=\int_{0}^\infty\mathrm{d}z\left[\rho(z)-\rho_b\right]$ denotes the surface excess, the surface tension of a repulsive interface grows with $\rho_b$ whereas for an attractive surface, $\sigma_e$ is a decreasing function of $\rho_b$. The former case, associated with the primitive model of ions at the water-air interface~\cite{wagner} is illustrated in Fig.~\ref{SurTen1}.a. The ionic depletion induced mainly by image-charge repulsion leads to an almost linearly increasing surface tension with $\rho_b$. The first important point to note in this figure is the reduction of $\sigma_e$ with ion size. This feature is particularly noticeable for sizes $a_i=2.5$ and 3 {\AA}. The decrease of $\sigma_e$ with increasing $a_i$ is due to the enhancement of the surface wetting of ions induced by core collisions (see also Fig.~\ref{figCHvsn}). Moreover, one notices that although $\sigma_e$ keeps its monotonic shape for $a_i=2.5$ {\AA} within the considered submolar concentration range, the curve for the largest ion size $a_i=3$ {\AA} exhibits a maximum at $\rho_b\simeq0.7$ M and decreases beyond this value. This reversal is also explained by the enhancement of excluded volume effects at large enough bulk concentrations and ion sizes, as confirmed by the comparison of Figs.~\ref{figCHvsn}.a and b.

We also computed ion densities and the corresponding PMFs in the presence of a Stern layer with $a_w=a_i$. The result is displayed in Fig.~\ref{PMF2}.a. A comparison of Figs.~\ref{figCHvsn} and Fig.~\ref{PMF2} shows that besides a sharp cut-off of ion concentrations at $z=a_w$, the most significant effect of the Stern layer is a decrease in the intensity of the repulsive image potential, which in turn leads to higher concentration peaks. This reduction of the image force is explained by the fact that the larger $a_w$ is, the larger is the separation between the central charge and the interface. Indeed, one can check in Eq.~(\ref{PMF1INT3}) that the amplitude of $V_{ci}(z)$ exponentially decreases with increasing $a_w$. A comparison of the surface tension curves in Figs.~\ref{SurTen1}.a and b confirms this observation : for a given ion size $a_i$, image forces are weakened by a finite $a_w$, which results in a stronger particle concentration at the interface and lowers the surface tension. At this point, we emphasize that our mapping between the ion size and the parameters of the repulsive Yukawa potential is not unique. A different choice will modify the range of the core interactions and lead to a quantitative change of the curves in Figs.~\ref{SurTen1}.a and b. However, this does not change our main conclusions here.

It was shown in this section that the consideration of core interactions between particles gives rise to a significant ionic adsorption at the neutral dielectric interface and the adsorption effect is amplified with the increase of the ion size or the bulk concentration. We also found that in agreement with Gibbs adsorption isotherm, this partial wetting lowers the surface tension of the primitive electrolyte model. We investigate in the next section the effect of core collisions on the exclusion of neutral and charged particles from slit pores.

\subsection{Slit pores}

In this section, we investigate the role of particle size on the exclusion of Yukawa particles from neutral slit pores, in contact with an external particle reservoir at the extremities. The pore geometry is depicted in Fig.~\ref{sketch}. In the neutral particle limit $q=0$ where Eq.~(\ref{eqvarSC1}) yields $\kappa_c=0$, partition coefficients and density profiles are obtained from the iterative solution of the variational equations~(\ref{eqvarEX2}) and~(\ref{eqvarSC2}) whereas for charged particles, the third  variational equation~(\ref{eqvarSC1}) for $\kappa_c$ should be included as well in the iterative algorithm.
\begin{figure}[t]
(a)\includegraphics[width=0.6\linewidth]{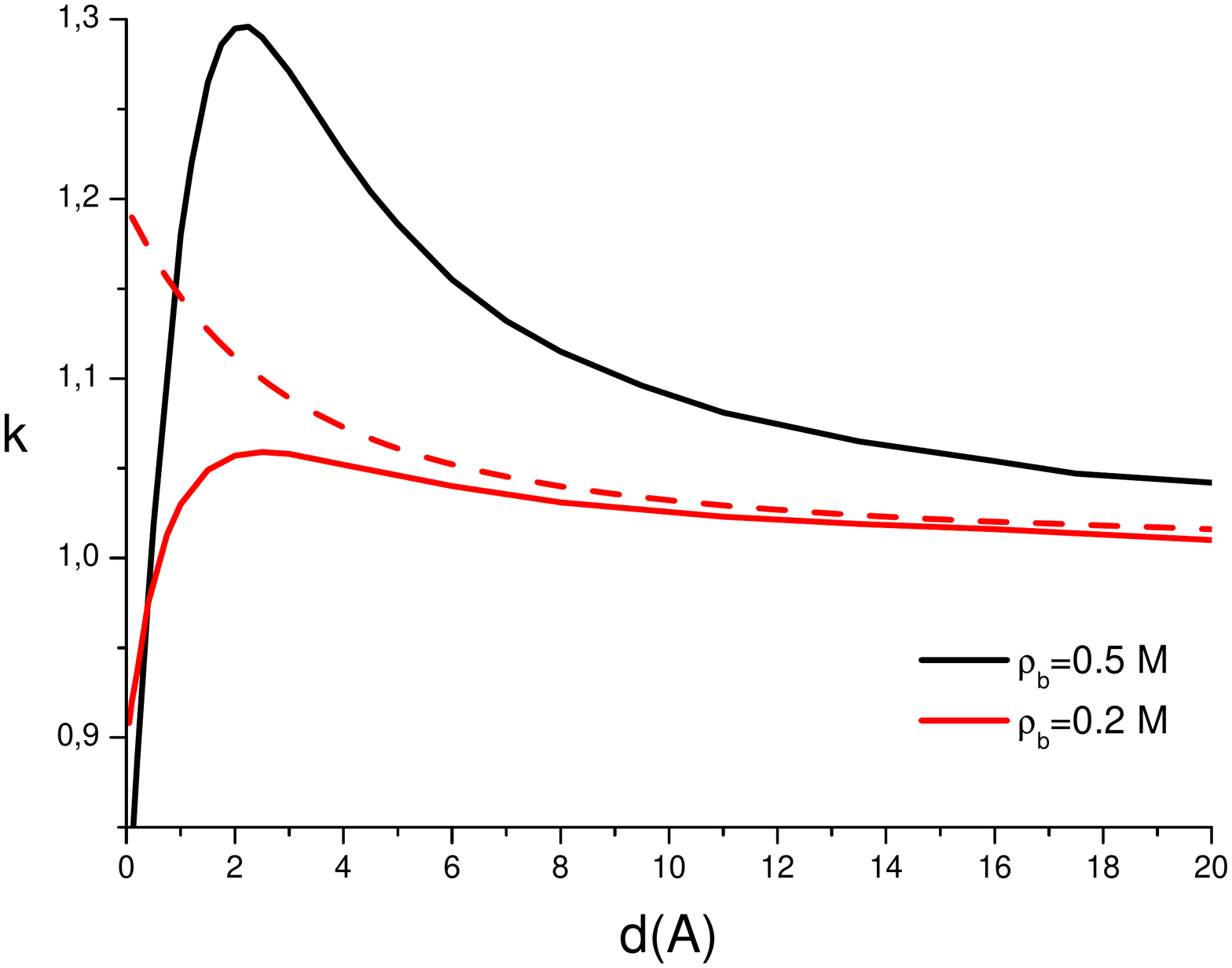}
(b)\includegraphics[width=0.6\linewidth]{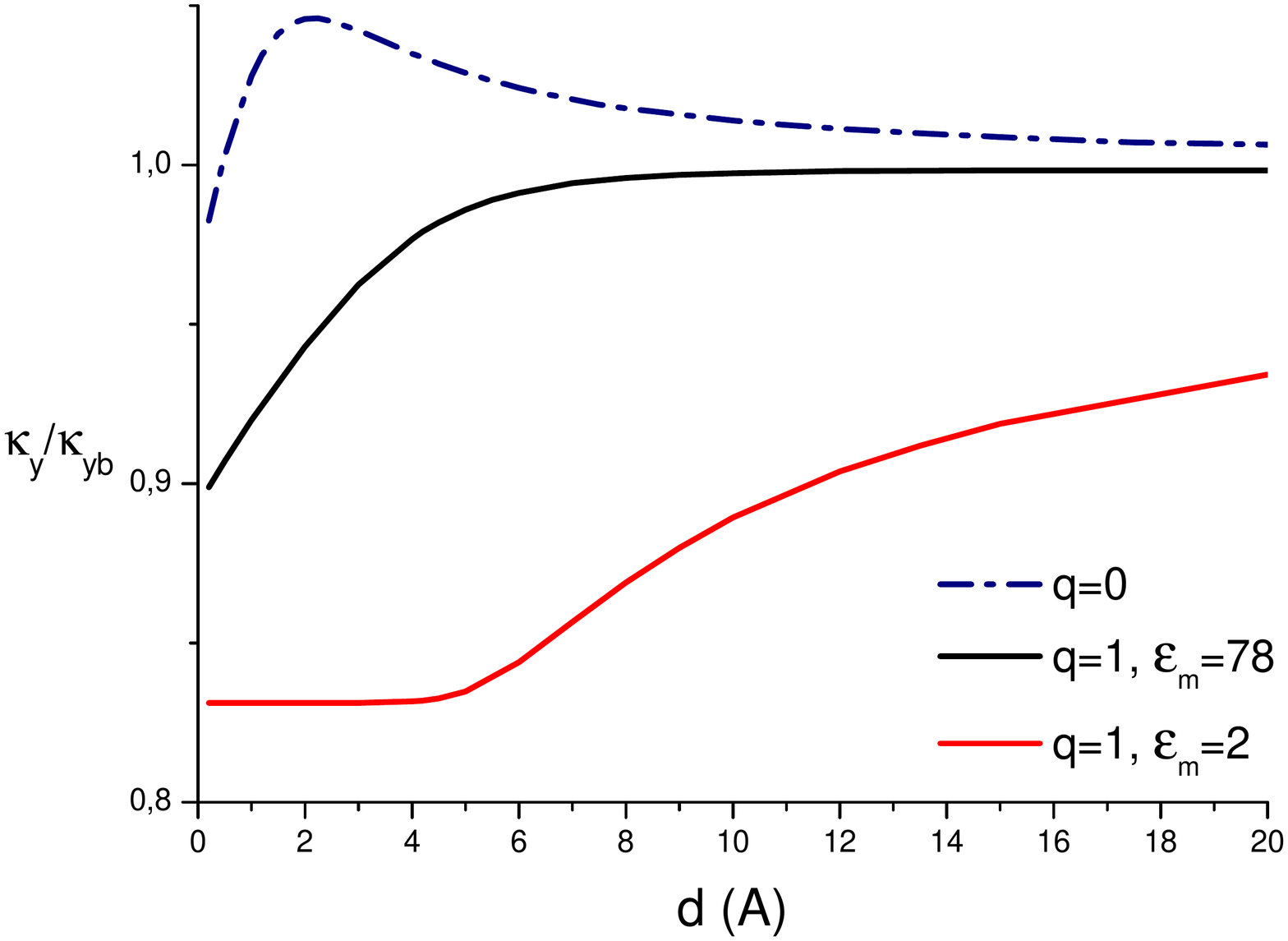}
\caption{(Color online) (a) Partition coefficient of neutral particles of size $a_i=2$ {\AA} against $d$ for $a_w=0$ at $\rho_b=0.2$ M (red curves) and $\rho_b=0.5$ M (black curve). The dashed line is the prediction of the MF theory and the solid lines display the variational result. (b) Variational inverse screening length $\kappa_y$ vs $d$ for neutral particles (blue dotted-dashed line) and monovalent ions in the cases $\e_m=\e_w$ (solid black line) and $\e_m=2$ (solid red line). $\rho_b=0.5$ M and the other model parameters are the same as in (a).}
\label{kSlitNeut}
\end{figure}

Fig.~\ref{kSlitNeut}.a displays the partition coefficient of neutral Yukawa particles of size $a_i=2$ {\AA} versus the pore size for $a_w=0$ and two different bulk concentrations. One notices that the MF theory predicts a pore density higher than the reservoir density. The particle adsorption into the pore is explained by core collisions that lead to an accumulation of particles at the boundaries that prevent their expansion. Since the confinement becomes stronger with decreasing pore size, the pore partition coefficient increases in a monotonic way. While the prediction of the variational calculation remains qualitatively similar for large pore sizes, the evolution of $k$ at small pore sizes significantly differs from the MF picture. At a characteristic pore thickness, $k$ reaches a maximum and then starts decreasing.

The reversal of the MF behavior originates from a competition between hard-core collisions and the pore-modified screening of these interactions. At the variational level, the latter effect is embodied in the integral term of the  potential Eq.~(\ref{PMFslit2}) that we will denote by $\delta V_y(z)$. In Section~\ref{1intNT} on single interfaces, it was shown that $\delta V_y(z)$ incorporates a solvation effect that pushes the particles towards the bulk region where Yukawa interactions can be screened more efficiently than next to the interface. This potential contains a similar effect in the slit pore. Indeed, $\delta V_y(z)$ excludes the particles from the pore. This feature can be understood by noting that particle penetration into the pore necessarily adds to the solvation and amplifies the intensity of $\delta V_y(z)$ (except at high densities where the screening of $\delta V_y(z)$ dominates the latter effect and $\delta V_y(z)$ begins to decrease with increasing $\rho_b$). Moreover, the reduction of the pore size also increases $\delta V_y(z)$ since the modification of the bulk screening is more significant in small pores. Consequently, $\delta V_y(z)$ becomes more repulsive with decreasing $d$ and at a characteristic pore size, its variation with $d$ dominates that of the attractive Yukawa potential $\psi_b-\psi_0(z)$ and $k$ begins to decrease. A comparison of the curves for $\rho_b=0.2$ M and $\rho_b=0.5$ shows that an increase of the reservoir density that positively adds to the amplitude of these opposing forces leads to a stronger competition between them and a higher peak for $k$. We also show in Fig.~\ref{kSlitNeut}.b the evolution of the variational inverse screening length $\kappa_y$ with the pore size. The inspection of the plot shows that $\kappa_y$ exhibits a similar shape to the partition coefficient, which is explained by the fact that $\kappa_y$ is an increasing function of the particle density in the pore (see Eq.~(\ref{eqvarSC2})). Finally, we note that in the case of neutral particles, adding a Stern layer simply reduces the accessible volume by $2Sa_w$. In other words, considering a finite $a_w$ is exactly equivalent to redefining the pore size according to $d'=d-2a_w$, which has the effect of shifting the curves in Fig.~\ref{kSlitNeut} towards larger pore sizes. This equivalence is valid for charged Yukawa particles exclusively in the case $\e_m=\e_w$.
\begin{figure}[t]
(a)\includegraphics[width=0.6\linewidth]{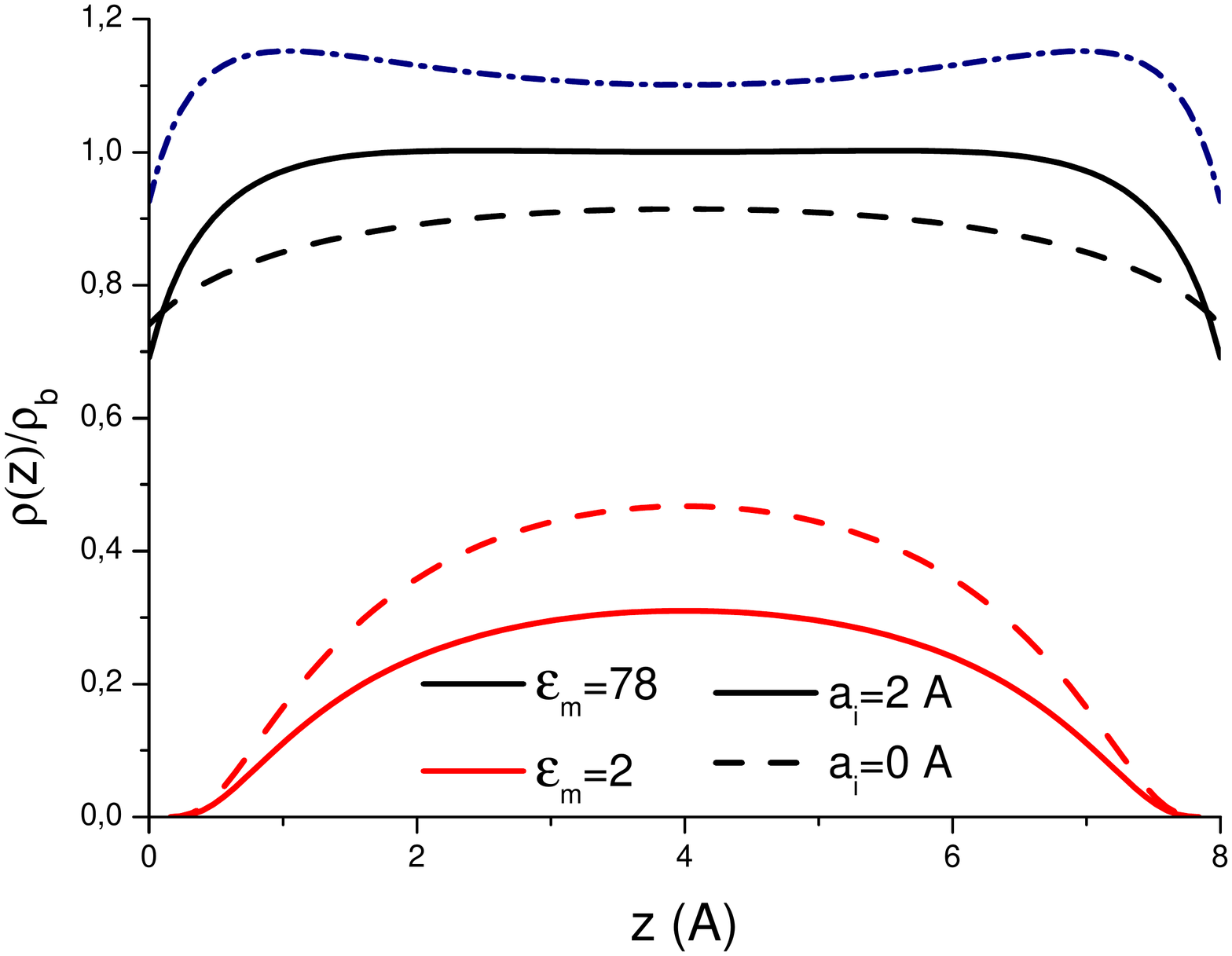}
(b)\includegraphics[width=0.6\linewidth]{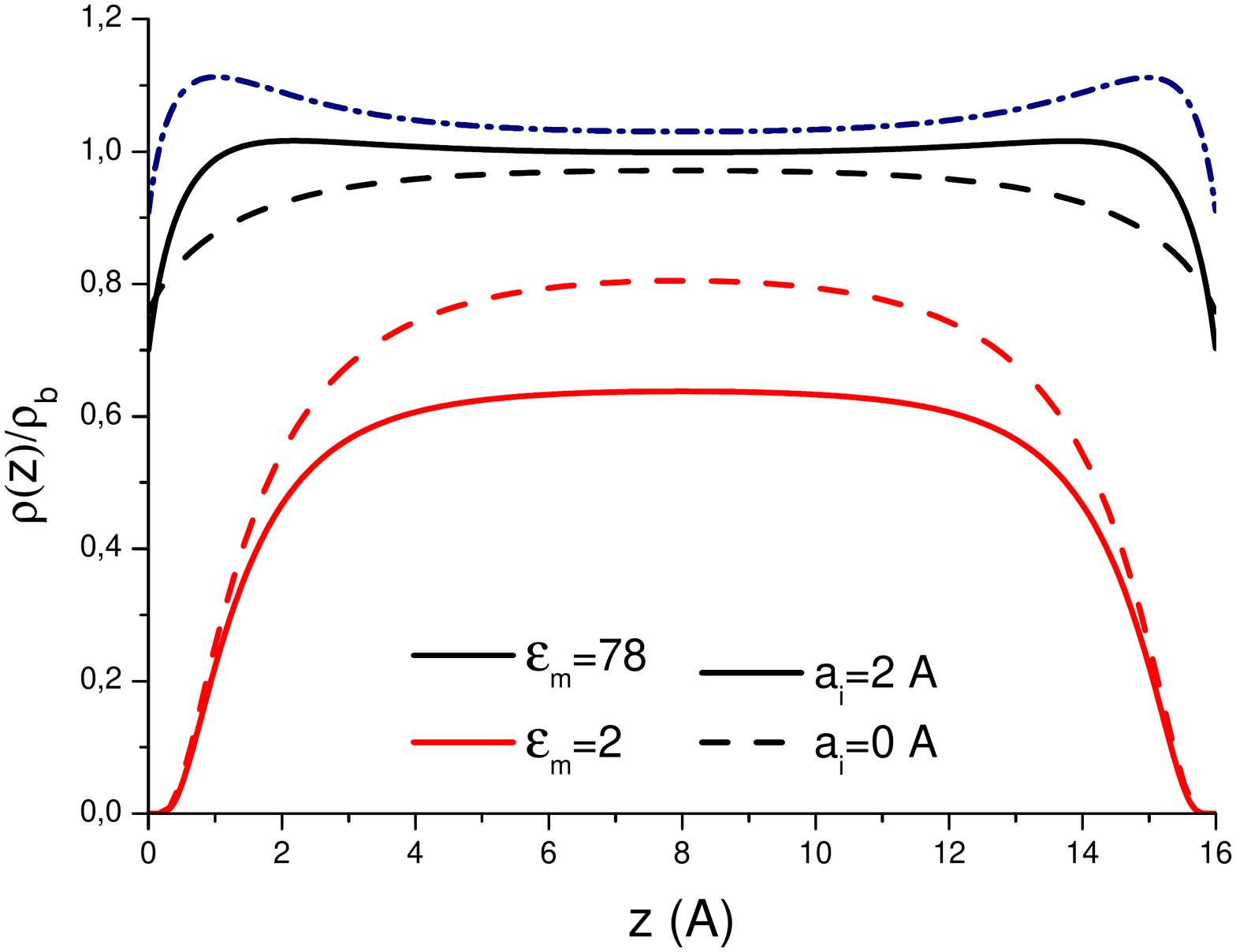}
\caption{(Color online)  Density profile of neutral Yukawa particles (blue dash-dotted line), monovalent ions without size (dashed lines) and with finite size $a_i=2$ {\AA} (solid lines) for (a) $d=8$ {\AA} and (b) $d=16$ {\AA}. Model parameters are $a_w=0$ {\AA}, $\rho_b=0.5$ M, $\e_w=78$ (black lines) and $\e_m=2$ (red lines).}
\label{rhoSlitCh}
\end{figure}

We compare in Fig.~\ref{rhoSlitCh} the density profile of neutral and charged Yukawa particles for $a_w=0$ and two different pore sizes, namely $d=8$ and $16$ {\AA}. First of all, one notices that the pore density profile of neutral Yukawa particles exhibits an oscillatory shape, with two concentration peaks that originate from the competition between the pore-modified screening of core interactions and core collisions that push the particles towards the pore walls. While the density profile of charged Yukawa particles is also oscillatory for $\e_m=78$, strong image forces that come into play in the case $\e_m=2$ smooth the density profile. Furthermore, as in the case of a single interface that was investigated in Sec.~\ref{1intNT}, the density profile of finite size ions is monotonic, and reaches its maximum value in the middle of the pore~\cite{hatlo,PRE}. This is explained by the fact that point-like ions are exclusively subject to repulsive image and electrostatic solvation forces, whose intensity reach their minimum value at $z=d/2$. More importantly, it is seen that even in the case of a vanishing dielectric discontinuity $\e_m=\e_w$, the local density of monovalent ions is lower than that of neutral particles. The reduction of the particle density is due to repulsive electrostatic solvation forces embodied in the potential $V_c(z)$ of Eq.~(\ref{PMFslit1}) that acts exclusively on particles of finite charge. Moreover, comparison of the curves for $\e_m=2$ in Fig.~\ref{rhoSlitCh}.a and Fig.~\ref{rhoSlitCh}.b shows that if the pore size is increased from $d=8$ {\AA} to $d=16$ {\AA}, the reduction of the intensity of image interactions leads to an extra ionic penetration into the pore. The underlying dielectric repulsion mechanism that is responsible for ionic exclusion from membrane nanopores was thoroughly investigated for point like ions in refs.~\cite{yarosch,hatlo,PRE}.
\begin{figure}[t]
(a)\includegraphics[width=0.6\linewidth]{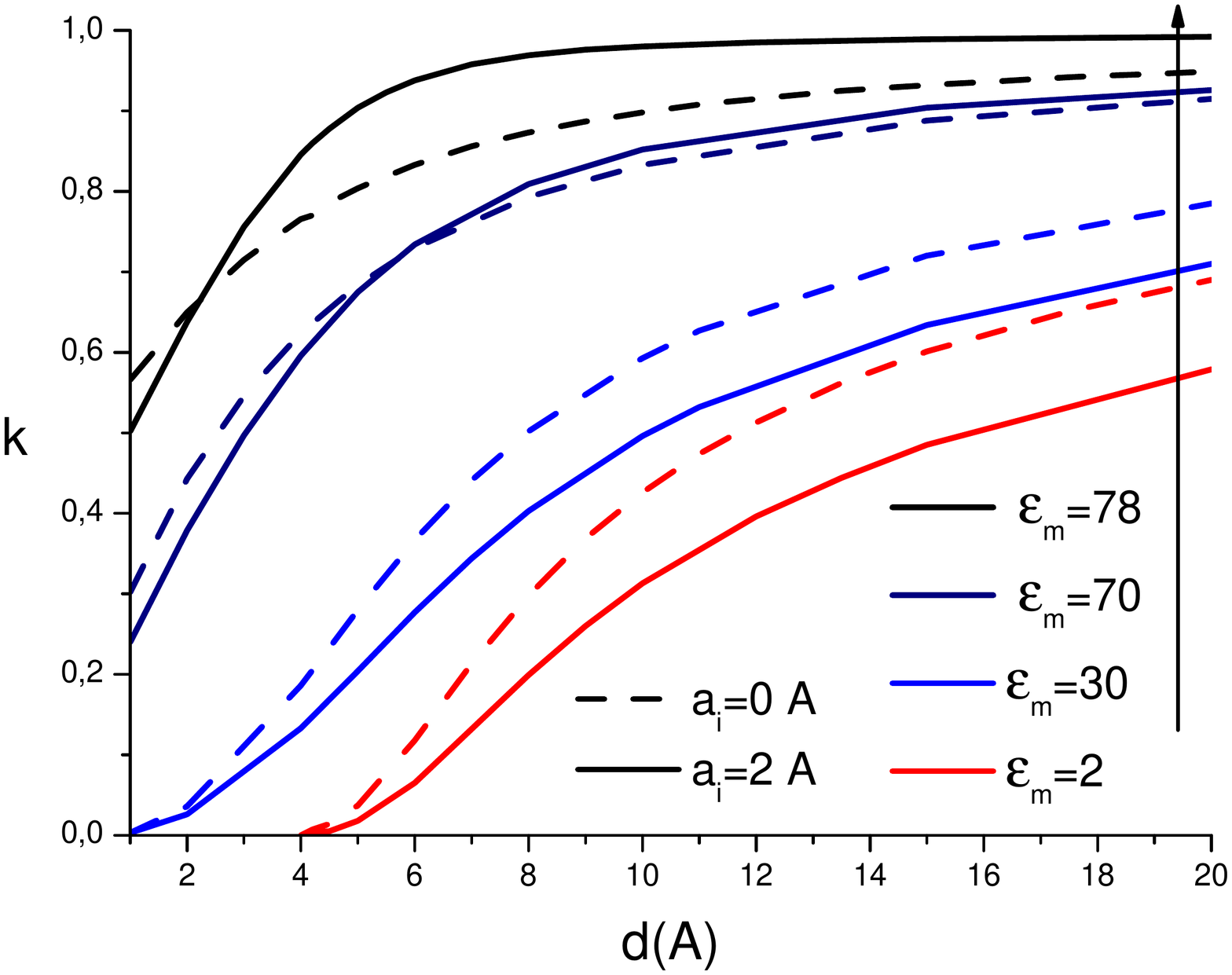}
(b)\includegraphics[width=0.6\linewidth]{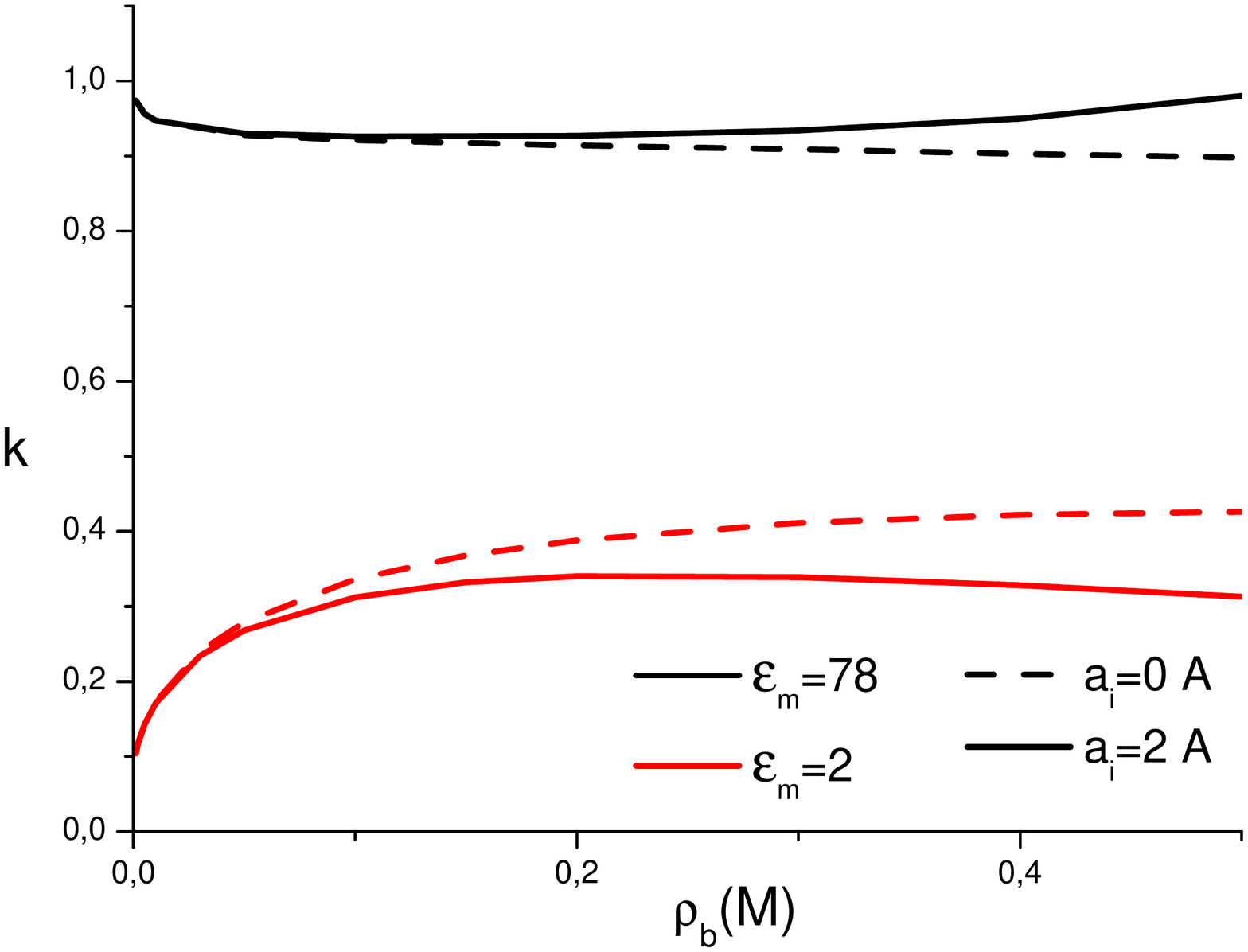}
\caption{(Color online) (a) Partition coefficient of monovalent ions without size (dashed lines) and of finite size $a_i=2$ {\AA} (solid lines) against the pore size $d$ for $a_w=0$ {\AA}, $\rho_b=0.5$ M, $\e_w=78$ and $\e_m=2$ to 78. (b) Partition coefficient against $\rho_b$ with $d=10$ {\AA} and the same model parameters as in (a).}
\label{kSlitCh1}
\end{figure}

We observe in Fig.~\ref{rhoSlitCh}.a and b that in the limit $\e_m=\e_w$, due to core collisions that push the particles into the pore, finite size ions have a larger density than point like ions. However, one also notices that the decrease of the membrane dielectric permittivity is accompanied with a highly interesting reversal. For $\e_m=2$, despite the core collisions that guide the finite size ions towards the pore, one finds that the local density of point like ions exceeds that of finite size ions. The same reversal effect is also illustrated in Fig.~\ref{kSlitCh1}.a which displays the ionic partition coefficient against the pore size for $\rho_b=0.5$ M and various values of $\e_m$. It is shown that in the case of a weak dielectric discontinuity and large pore sizes, the penetration of finite size ions is favored over that of point like ions, but for low values of $\e_m$ that correspond to strong image forces, the former experience a significantly stronger exclusion from the pore. This seemingly counter-intuitive effect is mainly due to the extra solvation energy barrier for the penetration of finite size ions into the pore. The energetic barrier is associated with the screening of core interactions, whose contribution to the particle density corresponds to the term $\ell_y(\kappa_{yb}-\kappa_y)$ of Eq.~(\ref{PMFslit2}) that vanishes for $a_i=0$ (or $\ell_y=0$). Indeed, a decrease of $d$ or $\e_m$, accompanied with an amplification of image-charge and solvation forces results in a reduction of the ion density and $\kappa_y$ (see Fig.~\ref{kSlitNeut}.b). As a result, when the ionic exclusion becomes strong enough, which occurs for small pores or low values of $\e_m$, the term $\ell_y(\kappa_{yb}-\kappa_y)$  reaches a large enough value to dominate the contribution from core collisions, i.e. the term $\psi(z)-\psi_b$ in Eq.~(\ref{locden}) and the density of finite size ions drop below that of point like ions.
\begin{figure}[t]
\label{kvsRHO}
\includegraphics[width=0.6\linewidth]{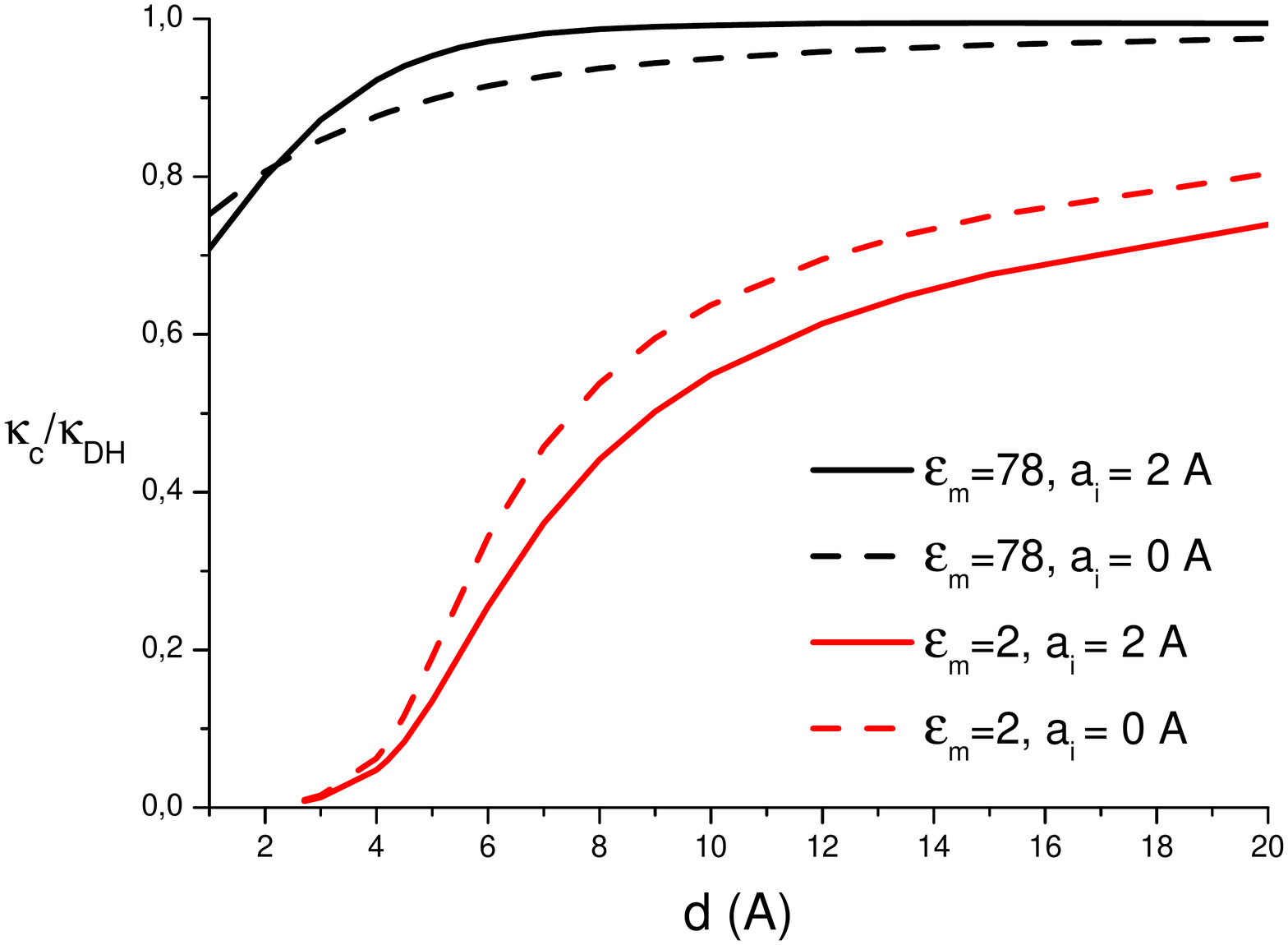}
\caption{(Color online)  Variational inverse screening length $\kappa_c$ against the pore size $d$  for monovalent ions without size (dashed lines) and with finite size $a_i=2$ {\AA} (solid lines). Model parameters are $a_w=0$ {\AA}, $\rho_b=0.5$ M, $\e_w=78$ (black lines) and $\e_m=2$ (dashed lines).}
\label{kappac}
\end{figure}

Fig.~\ref{kSlitCh1}.b displays the partition coefficient of finite size and point like ions against their bulk concentration. It is shown that the reversal of the facilitated penetration of large size particles with decreasing $\e_m$ extends over the whole concentration regime, i.e. this peculiarity also survives in the very dilute limit. However, since the decrease of the bulk concentration results in a reduction of excluded volume effects, the difference between the pore density of point-like and finite size ions monotonically decreases with their bulk density.

The evolution of variational inverse screening lengths for monovalent ions is illustrated as a function of the pore size in Fig.~\ref{kSlitNeut}.b and Fig.~\ref{kappac}. First of all, we notice in Fig.~\ref{kappac} that $\kappa_c$ also exhibits the reversal effect discussed above. Then, the inspection of Fig.~\ref{kSlitNeut}.b and Fig.~\ref{kappac} shows that as expected from the variational equations~(\ref{eqvarSC1}) and~(\ref{eqvarSC2}), $\kappa_c$ and $\kappa_y$ qualitatively follow the same trend as the partition coefficient. Namely, in the case $\e_m=\e_w$ where the ionic exclusion is not too strong, $\kappa_c$ and $\kappa_y$ monotonically decrease from their bulk values towards slightly lower values with decreasing $d$. In the opposite limit $\e_m=2$ that corresponds to strong image forces, at a characteristic pore thickness below which the salt rejection is almost total ($\lan\rho(z)/\rho_b\ran\simeq0$ for $d\simeq 4$ {\AA}), the system reaches the dilute limit, and one gets $\kappa_y\simeq b$ and $\kappa_c\simeq0$.

\begin{figure}[t]
\label{kvsRHO}
\includegraphics[width=0.6\linewidth]{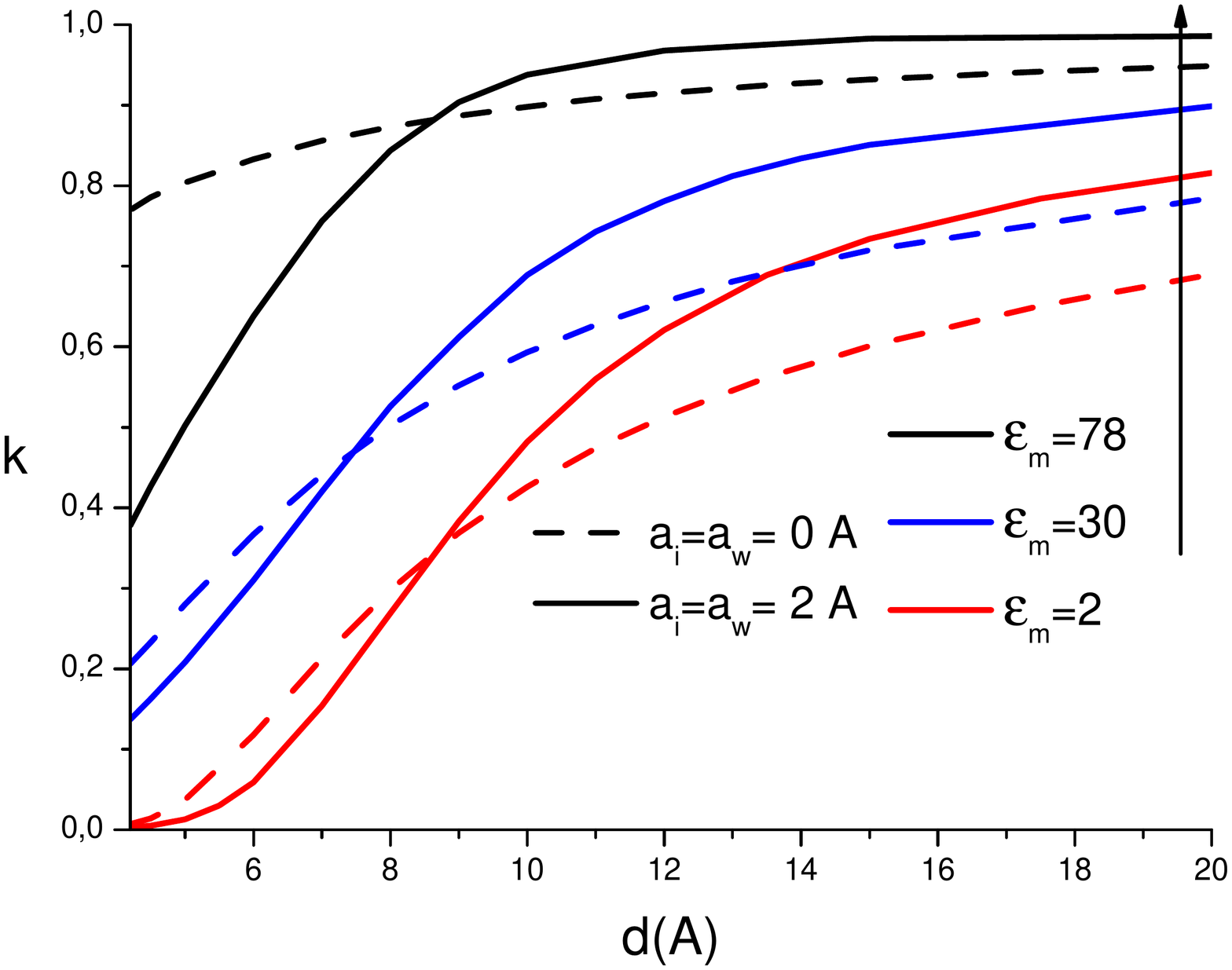}
\caption{(Color online)  Partition coefficient of monovalent ions against the pore size $d$ with a finite Stern layer of thickness $a_w=a_i$. Model parameters are the same as in Fig.~\ref{kSlitCh1}.a.}
\label{kSlitCh2}
\end{figure}

We also plot in Fig.~\ref{kSlitCh2} ionic partition coefficients as a function of the pore size with a finite Stern layer $a_w=a_i$. In addition to an effective reduction of the pore volume accessible to ions, it was shown in Sec.~\ref{1intNT} that a finite $a_w$ increases the distance between the interface and the central charge, which results in an overall reduction of the image repulsion. Thus, one expects these two opposing effects to shift the curves in Fig.~\ref{kSlitCh1}.a towards larger pore sizes along the abscissa axis and also to larger densities along the ordinate axis. As a result, Fig.~\ref{kSlitCh2} shows that for any membrane dielectric permittivity satisfying $\e_m\leq\e_w$, ionic penetration into the pore is characterized by a stronger rejection of finite size ions at small $d$ and their favored penetration over point like ions for large $d$.

\section{Conclusion}

In this article, we have studied a field-theoretic model that considers electrostatic and core interactions on an equal footing in order to investigate the role played by excluded volume effects on the partition of ions at dielectric interfaces and ionic selectivity of slit nanopores. To this end, we have developed a variational calculation scheme that allows to go beyond the MF regime and couples in a consistent way pore modified core interactions, steric effects, electrostatic solvation and image-charge forces, and electrostatic potential induced by a surface charge distribution.

In the first part of the work, we considered a bulk liquid composed of charged Yukawa particles. Using a general variational ansatz, it was shown that at the first order variational level, short range core interactions between two test ions experience a further screening from the surrounding particles but electrostatic interactions between them are not modified by core collisions.

The second part of the article was devoted to the configuration of ions at simple interfaces. Density profiles for neutral Yukawa particles obtained from the MF limit of the theory and the variational formalism, and also from a recent RPA method were compared with MC simulations. We showed that in the submolar concentration regime, the MF theory presents a reasonable agreement with simulation results, but slightly overestimates particle concentrations in the proximity of the hard wall. It was shown that the variational theory, which includes the correlation effects associated with the modification of the screening of core interactions by the interface gives a better agreement with simulation results. We also observed a deviation of the variational prediction from MC results very close to the interface, namely for $z\lesssim 0.5$ {\AA}. This discrepancy is probably be due to the simple choice of a uniform variational screening length for core interactions that does not account for the variation of the particle density close to the wall. It was also shown that the RPA method overestimates concentration profiles over the whole interfacial area. In the case of charged particles in contact with a dielectric interface, the competition between core collisions and repulsive dielectric forces yields concentration peaks at the interface and the corresponding ionic adsorption is amplified with increasing ion size or bulk electrolyte concentration. We computed an integral expression for the surface tension that accounts for excluded volume effects and also a Stern layer, induced by the finite ion size. In agreement with the isothermal Gibbs adsorption equation, we showed that an increase of the ion size that yields a stronger wetting on the wall is also accompanied with a reduction of the surface tension.

In the third part, we solved the variational equations in a neutral slit pore. We first treated the case of neutral particles to show that excluded volume effects associated with core collisions within the bulk reservoir result in a net particle adsorption into the pore. In the absence of a dielectric  discontinuity, we found that due to the same core collisions that drive the particles into the pore, the density of ions increases with their size. However, the presence of a dielectric discontinuity reverse this picture : it was shown that the ionic selectivity of pores characterized with strong image forces increases with the ion size. The reversal of the pore selectivity originates from a complex coupling with the solvation energy associated with core interactions and image forces, and survives in the limit of dilute electrolytes. This interesting peculiarity calls for experimental confirmation. The present pore model includes as well a Stern layer $a_w$  associated with the surface affinity of ions. Besides a reduction of the volume accessible to ions, a finite $a_w$ decreases the intensity of repulsive image forces. As a result, small pores of subnanometric size exhibit a stronger selectivity for large ions but nanometric pores exclude small ions more efficiently than large ions.

In this article, we have investigated exclusively the case of neutral planar interfaces. However, the variational equations~(\ref{eqvarEX1})-(\ref{eqvarSC2}) are derived for the general case of charged interfaces. The role played by excluded volume effects on ionic exclusion from charged pores will be presented in future work. We note that the relations Eqs.~(\ref{eqvarEX1})-(\ref{eqvarSC2}) can also treat the case of curved dielectric bodies such as spherical colloids or proteins of cylindrical geometry. The consideration of curved geometries deserve further investigation, since the curvature is known to come into play in various problems of biological interest, such as ionic exclusion from membrane nanopores of cylindrical geometry~\cite{PRL,JCP} or counterion condensation onto charged proteins~\cite{Bor}. Furthermore, the derived variational equations generalize the approximative self-consistent equations used in nanofiltration theories~\cite{yarosch,Szymczyk,YarII} to the case of ions associated with important core interactions and hydrodynamic forces deserve to be included in the model for applications related to ionic transport phenomenon~\cite{Ge}.

The present model has of course several limitations. First of all, our variational calculation scheme is based on a generalized Onsager-Samaras approximation that assumes a uniform screening parameter for electrostatic and core interactions. As mentioned above, this simple choice might be responsible for the deviation of the variational result from simulation data in the close neighborhood of the interface. A more sophisticated variational choice, such as a spatially varying screening length is expected to correct the observed discrepancy. We currently investigate this point. Moreover, the variational equations were derived from a first order variational free energy, which limited our investigation to the submolar concentration regime. Corrections from a second order cumulant expansion might be necessary to consider higher concentration regimes.

We finally note that in the case of ions confined into closed geometries characterized by a dielectric discontinuity, the existence of an infinite number of electrostatic images considerably complicates reliable numerical computations such as MC simulations. At the present, we are not aware of any simulation data for charged Yukawa liquids in contact with dielectric interfaces. The generalization of recent MC algorithms developed for confined Coulomb liquids to the case of charged Yukawa liquids will able us to determine in the future the validity domain of the present formalism~\cite{Arnold1,Arnold2}.
\\
\acknowledgements  Sahin Buyukdagli thanks John Palmeri, David S Dean and Manoel Manghi for helpful
discussions. This work has been in part supported by The Academy of Finland through its COMP CoE grant.

\smallskip

\appendix
\section{Boundary conditions for the external fields}
\label{BounCon}

This Appendix is devoted to the derivation of the boundary conditions that come into play when solving Eqs.~(\ref{MFphi}) and~(\ref{MFpsi}) or Eqs.~(\ref{MFphi}) and~(\ref{MFpsi}). In the case of a charged HC liquid confined between two planar surfaces located at $z=0$ and $z=d$, with surface charge $\sigma(z)=\sigma_s\left[\delta(z)+\delta(z-d)\right]$, the first step to derive the boundary conditions consists in integrating Eqs.~(\ref{MFphi}) and~(\ref{MFpsi}) over the intervals $z\in[0^-,0^+]$  and $z\in[a_w^-,a_w^+]$. One obtains
\bea\label{boundphi0}
&&\left.\frac{d\phi}{dz}\right|_{z=0^+}=\left.\frac{d\phi}{dz}\right|_{z=0^-}-4\pi q\ell_w\sigma_s\\
\label{boundphia}
&&\left.\frac{d\phi}{dz}\right|_{z=a^+}=\left.\frac{d\phi}{dz}\right|_{z=a_w^-}\\
\label{boundpsi0}
&&\left.\frac{d\psi}{dz}\right|_{z=0^+}=\left.\frac{d\psi}{dz}\right|_{z=0^-}\\
\label{boundpsia}
&&\left.\frac{d\psi}{dz}\right|_{z=a^+}=\left.\frac{d\psi}{dz}\right|_{z=a_w^-}.
\eea
Furthermore, in the particle free region $z<a_w$, $\rho_b=0$ and Eqs.~(\ref{MFphi}), (\ref{MFpsi}) reduce to
\bea
\label{MFphi2}
&&\frac{d^2\phi}{dz^2}=0\\
\label{MFpsi2}
&&\frac{d^2\psi}{dz^2}-b^2\psi=0.
\eea
The solution of Eqs.~(\ref{MFphi2}) and~(\ref{MFpsi2}) that satisfy the boundary conditions $\phi(z\to-\infty)=0$ and $\psi(z\to-\infty)=0$ is
\bea\label{eq1}
&&\phi(z)=0\hspace{0.5mm}\mbox{,}\hspace{4mm}z<0\\
\label{eq2}
&&\phi(z)=c_1+c_2z\hspace{0.5mm}\mbox{,}\hspace{4mm}0<z<a_w\\
\label{eq3}
&&\psi(z)=c_3e^{bz}\hspace{0.5mm}\mbox{,}\hspace{4mm}z<a_w
\eea
where $c_1,c_2,c_3$ are constants of integration. Deriving Eq.~(\ref{eq3}), we assumed the continuity of the Yukawa field, i.e. $\psi(0^+)=\psi(0^-)$. According to Eq.~(\ref{eq1}), the derivative of the electrostatic potential in the substrate part $z<0$ vanishes, $d\phi(z)/dz=0$, which yields with Eq.~(\ref{boundphi0}) the Gauss law $\left.d\phi(z)/dz\right|_{0^+}=-4\pi q\ell_w\sigma_s$. Thus Eq.~(\ref{eq2}) becomes
\be\label{eq22}
\phi(z)=c_1-4\pi q\ell_w\sigma_sz\hspace{0.5mm}\mbox{,}\hspace{4mm}0<z<a_w.
\ee
By injecting Eq.~(\ref{eq22}) into Eq.~(\ref{boundphia}) and Eq.~(\ref{eq3}) into Eq.~(\ref{boundpsia}) and imposing the continuity of the Yukawa potential $\psi(z=a_w^-)=\psi(z=a_w^+)$, we finally obtain the boundary conditions Eqs.~(\ref{boundphi2})-(\ref{boundpsi2}) for the derivatives of the potentials at $z=a_w$. Following the same steps as above for the interface located at $z=d-a_w$, one gets the additional boundary conditions Eqs.~(\ref{boundphi3})-(\ref{boundpsi3}).

\section{Variational Yukawa and electrostatic Green's functions}
\label{appendixVarFr}
In this Appendix, we explain the derivation of the variational HC potential $w_0(\br,\br')$ and the electrostatic potential $v_0(\br,\br')$ for the slit pore geometry depicted in Fig.~\ref{sketch}. These potentials are obtained by inverting the relations~(\ref{DH1}) and (\ref{DH2}), i.e. by solving a generalized DH equation
\be\label{DH3}
\left[-\nabla(\epsilon(\br)\nabla)+\epsilon(\br)\kappa^2(\br)\right]U(\br,\br')=\lambda\delta(\br-\br')
\ee
where the free parameter $\lambda$ will be adjusted later in order to recover $v_0(\br,\br')$ and $w_0(\br,\br')$ from the kernel $U(\br,\br')$. The piecewise dielectric permittivity and the inverse screening length, respectively, are given by
\bea\label{pwe}
\e(z)&=&\e_>\theta(z)\theta(d-z)+\e_<[\theta(-z)+\theta(z-d)]\hspace{0.5cm}\\
\label{pwk}
\kappa_y(z)&=&\kappa_<[\theta(a_w-z)+\theta(z+a_w-d)]+\kappa_>\theta(z-a_w)\theta(d-a_w-z).
\eea
By injecting into Eq.~(\ref{DH3}) the Fourier expansion of the kernel
\be\label{kernel1}
U(\br,\br')=\int\frac{\mathrm{d}^2\bk}{4\pi^2}e^{i\bk\cdot(\br-\br')} \tilde U(z,z',k),
\ee
one gets
\be\label{DH32}
\left[-\partial_z\epsilon(z)\partial_z+\epsilon(z)\left(k^2+\kappa^2(z)\right)\right]\tilde U(z,z',k)=\lambda\delta(z-z').
\ee
The homogeneous solution to this equation is of the form $U=C_+e^{\sqrt{k^2+\kappa^2}z}+C_-e^{-\sqrt{k^2+\kappa^2}z}$. The coefficients $C_\pm$ for each layer are found by imposing the finiteness of the kernel $\lim_{z\to\pm\infty} \tilde U(z)=0$ and the boundary conditions
\bea
&&\tilde U\left(z=\Sigma_-\right)=\tilde U\left(z=\Sigma_+\right)\\
&&\e\left(z=\Sigma_-\right)\left.\frac{\partial\tilde U}{\partial z}\right|_{z=\Sigma_-}=\e\left(z=\Sigma_+\right)\left.\frac{\partial \tilde U} {\partial z}\right|_{z=\Sigma_+}\\
&&\left.\frac{\partial\tilde U}{\partial z}\right|_{z=z'_+}-\left.\frac{ \partial \tilde U}{\partial z}\right|_{z=z'_-}=-\lambda/\e_>
\eea
where $\Sigma$ denotes the position of the surfaces at $z=0,a_w,d-a_w,d$ and we also introduced the notation $\tilde U\left(\Sigma_\pm\right)\equiv\lim_{\epsilon\to0}\tilde U\left(\Sigma\pm\epsilon\right)$. The kernel reads for $0<z,z'<d$
\be\label{kernel0}
U(\br,\br')=\frac{\lambda}{4\pi\e_>}\frac{e^{-\kappa_>|\br-\br'|}}{|\br-\br'|}+\delta U(\br,\br')
\ee
with the anisotropic part
\bea\label{kernel1}
&&\delta U(\br,\br')=\int\frac{\mathrm{d}^2\bk}{4\pi^2}e^{i\bk\cdot(\br-\br')}\delta \tilde U(z,z',k)\\
\label{kernel2}
&&\delta \tilde U(z,z',k)=\frac{\lambda}{2\e_>\rho_>}\frac{\Delta}{1-\Delta^2e^{-2\rho_>(d-2a_w)}}\times\left[e^{-\rho_>(z+z'-2a_w)}
+e^{\rho_>(z+z'+2a_w-2d)}+2\Delta e^{-2\rho_>(d-2a_w)}\cosh\left(\rho_>|z-z'|\right)\right]\nonumber\\
\eea
where we have defined
\be
\Delta=\frac{\rho_>-\eta\rho_<}{\rho_>+\eta\rho_<},\hspace{5mm}\eta=\frac{1-\Delta_0e^{-2\rho_<a_w}}{1+\Delta_0e^{-2\rho_<a_w}}
\ee
and $\rho_\lessgtr=\sqrt{k^2+\kappa_\lessgtr^2}$,  $\Delta_0=(\e_>-\e_<)/(\e_>+\e_<)$. By comparing the relations~(\ref{DH1}) and (\ref{DH2}) with Eq.~(\ref{DH3}), the variational potentials $v_0(\br,\br')$ and $w_0(\br,\br')$ directly follow from the solution~(\ref{kernel2}) as explained below.

First, by setting $\kappa_<=0$, $\kappa_>=\kappa_c$, $\e_<=\e_m$, $\e_>=\e_w$, $\lambda=4\pi\e_>\ell_w$ and defining $\rho_c=\sqrt{\kappa_c^2+k^2}$,
\be
\Delta_c=\frac{\rho_c-\eta_c k}{\rho_c+\eta_ck} ,\hspace{5mm}\eta_c=\frac{1-\Delta_0e^{-2ka_w}}{1+\Delta_0e^{-2ka_w}},
\ee
and $\Delta_0=(\e_w-\e_m)/(\e_w+\e_m)$, the variational electrostatic potential is obtained from Eq.~(\ref{kernel0}) in the form
\bea
v_0(\br,\br')=\ell_w\frac{e^{-\kappa_c|\br-\br'|}}{|\br-\br'|}+\delta v_0(\br,\br')
\eea
where
\bea\label{kernel3}
&&\delta v_0(\br,\br')=\int\frac{\mathrm{d}^2\bk}{4\pi^2}e^{i\bk\cdot(\br-\br')}\delta \tilde v_0(z,z',k)\\
\label{kernel4}
&&\delta \tilde v_0(z,z',k)=\frac{2\pi\ell_w
}{\rho_c}\frac{\Delta_c}{1-\Delta_c^2e^{-2\rho_c(d-2a_w)}}\left[e^{-\rho_c(z+z'-2a_w)}
+e^{\rho_c(z+z'+2a_w-2d)}+2\Delta_c e^{-2\rho_>(d-2a_w)}\cosh\left(\rho_>|z-z'|\right)\right]\nonumber.\\
\eea

Second, by setting $\kappa_<=b$, $\kappa_>=\kappa_y$, $\e_<=\e_>$, $\lambda=4\pi\e_>\ell_y$ and defining $\rho_m=\sqrt{b^2+k^2}$, $\rho_y=\sqrt{\kappa_y^2+k^2}$ and
\be
\Delta_y=\frac{\rho_y-\rho_m} {\rho_y+\rho_m},
\ee
the Yukawa potential follows from Eq.~(\ref{kernel0}) as
\bea\label{kernel4.5}
w_0(\br,\br')=\ell_y\frac{e^{-\kappa_y|\br-\br'|}}{|\br-\br'|}+\delta w_0(\br,\br')
\eea
where
\bea\label{kernel5}
&&\delta w_0(\br,\br')=\int\frac{\mathrm{d}^2\bk}{4\pi^2}e^{i\bk\cdot(\br-\br')}\delta \tilde w_0(z,z',k)\\
\label{kernel6}
&&\delta \tilde w_0(z,z',k)=\frac{2\pi\ell_y}{\rho_y}\frac{\Delta_y}{1-\Delta_y^2e^{-2\rho_y(d-2a_w)}}\left[e^{-\rho_y(z+z'-2a_w)}
+e^{\rho_y(z+z'+2a_w-2d)}+2\Delta_y e^{-2\rho_y(d-2a_w)}\cosh\left(\rho_y|z-z'|\right)\right]\nonumber.\\
\eea

\section{Computation of the surface tension with the charging procedure}
\label{SurTenTen}

We explain in this Appendix the computation of the surface tension $\sigma_e$ associated with a neutral interface. The surface tension is defined as the excess Grand potential,
\be\label{surten1}
\sigma_e=\frac{k_BT}{S}\left[\Omega_v-\Omega_v(\rho_b=0)-\Omega_1\right]
\ee
where the Grand potential of the bulk system given by Eq.~(\ref{BulkGr}) reads
\be\label{PotBl}
\frac{\Omega_1}{SL} =\frac{1}{24\pi}\left[\kappa_v^3+(\kappa_{yb}-b)(\kappa_{yb}^2+\kappa_{yb}b-2b^2)\right]
-\frac{b^2\psi_b^2}{8\pi\ell_y}-2\rho_b.
\ee
In the above equation, $L$ stands for the length of the system. We also note that in Eq.~(\ref{surten1}), the subtracted quantity $\Omega_{0\varphi}(\rho_b=0)$ corresponds to the vdW level surface tension of the pure water.

The derivation of the vdW part of the surface tension $\Omega_0$ with the charging procedure was explained in Ref.~\cite{netzvdw} for the DH theory of slit pores. Using a general kernel as in Appendix~\ref{appendixVarFr}, the Coulomb and Yukawa contributions Eqs.~(\ref{GausCoPhi}) and~(\ref{GausCoPsi}) will be derived with a single calculation. The charging method consists in reexpressing the excess van der Waals energy in the form of an integral over an auxiliary charging parameter $\xi$,
\bea\label{ex1}
\Delta\Omega_{0\varphi}\equiv\Omega_{0\varphi}-\Omega_{0\varphi}(\rho_b=0)&=&-\ln
\frac{\int\mathcal{D}\varphi\;e^{-\frac{1}{2}\int_{\br,\br'}\varphi(\br)U^{-1}\left(\br,\br';\kappa(z)\right)\varphi(\br')}}
{\int\mathcal{D}\varphi\;e^{-\frac{1}{2}\int_{\br,\br'}\varphi(\br)U^{-1}\left(\br,\br';\kappa_1\right)\varphi(\br')}}\\
&=&-\int_0^1\mathrm{d}\xi\frac{\mathrm{d}}{\mathrm{d}\xi}\ln\int\mathcal{D}\varphi\;e^{-\int\frac{\mathrm{d}\br}{2\lambda}\e(\br)
\left[(\nabla\varphi)^2+\kappa_\xi^2(\br)\right]}\nonumber
\eea
where the operator $U^{-1}\left(\br,\br';\kappa(z)\right)$ is defined by Eq.~(\ref{DH3}), whose solution for a single interface geometry follows from Eq.~(\ref{kernel1}) in the limit $d\to\infty$. In the same limit, the piecewise functions $\epsilon(z)$ and $\kappa(z)$ defined in Eqs.~(\ref{pwe}) and~(\ref{pwk}) become $\e(z)=\e_>\theta(z)+\e_<\theta(-z)$ and $\kappa_y(z)=\kappa_<\theta(a_w-z)+\kappa_>\theta(z-a_w)$. As in Appendix~\ref{appendixVarFr}, the free parameter $\lambda$ will be adjusted later and the fluctuating potential $\varphi$ will be identified with $\phi$ or $\psi$ at the end of the calculation in order to recover $\Omega_{0\phi}$ and $\Omega_{0\psi}$ from $\Omega_{0\varphi}$. In Eq.~(\ref{ex1}), we have also introduced the function $\kappa_\xi^2(\br)=\kappa_<^2+\xi\left[\kappa(\br)^2-\kappa_<^2\right]$. Evaluating the derivative with respect to $\xi$ in Eq.~(\ref{ex1}), one obtains
\be
\Delta\Omega_{0\varphi}=S\frac{\e_>(\kappa_>^2-\kappa_<^2)}{2\lambda}\int_0^1\mathrm{d}\xi\int_{a_w}^\infty\mathrm{d}z U(\br,\br;\kappa_\xi)
\ee
Carrying out the integral over $z$ and substituting as in  Appendix~\ref{appendixVarFr} $\lambda$, $\kappa_\lessgtr$ and $\e_\lessgtr$ by their values associated with the Yukawa and Coulomb contributions, one recovers $\Delta\Omega_{0\psi}$ and $\Delta\Omega_{0\phi}$ from $\Delta\Omega_{0\varphi}$. Injecting $\Omega_v$ of Eq.~(\ref{grandPot}) with these vdW-level contributions and the bulk Potential Eq.~(\ref{PotBl}) into the relation~(\ref{surten1}), one finally obtains the surface tension Eq.~(\ref{surten}).

\section{Monte Carlo simulations}
\label{mcdetails}
We present in this Appendix the details of the canonical MC simulations. All simulation results presented in section \ref{1intNT} were obtained by performing standard Monte Carlo simulations \cite{simulationbooks} in a slab geometry of size $l_x=l_y=20\rho^{-1/3}_b$ and $l_z=50\rho^{-1/3}_b$, with impenetrable walls located at $z=0$ and $z=l_z$. Periodic boundary conditions in the $x$ and $y$ directions were used. The value of $l_z$ that we considered is large enough so that the box geometry is equivalent to the single interface case presented in section \ref{1intNT}. During the simulations, the total number of particles was fixed at $N_p=20000$, with the corresponding particle density $\bar \rho_{\textrm{MC}}=N_p/(l_xl_yl_z)=\rho_b$. The average acceptance rate during the simulations was approximately $0.80$. For each set of parameters, the system is initialized with a random configuration and first equilibrated over $50000$ Monte Carlo steps. After equilibration, the density profiles are obtained according to
\begin{equation}
\rho(z)=\frac{1}{dz}\frac{1}{l_xl_y}\lan\sum_{\{z\leq z_i<z+dz\}_i} 1\ran,
\end{equation}
where $z_i$ is the coordinate of the particle $i$ along $z$ axis, $dz=\rho^{-1/3}_b/50$ is the width of a transversal bin and $<\cdot>$ denotes the ensemble average. For all cases, the average was obtained from 20 independent runs with $5000$ measurements taken at $20$ Monte Carlo steps interval for each run. The density profiles were obtained from an average over 20 independent runs, with $5000$ measurements taken at intervals of $20$ Monte Carlo steps for each run.

\end{document}